\documentclass[journal]{IEEEtran}
\IEEEoverridecommandlockouts
\usepackage[dvips]{graphicx}
\usepackage[english]{babel}
\usepackage{algorithm}
\usepackage{algorithmic}
\usepackage{amsfonts}
\usepackage{booktabs}
\usepackage{multirow}
\usepackage{upgreek}
\usepackage{amsmath}
\usepackage{amssymb}
\usepackage{cite}
\usepackage{array}
\usepackage{url}
\usepackage{caption}
\usepackage{bigstrut}
\usepackage{subfigure}
\usepackage{multirow}
\usepackage{balance}
\usepackage{float}

\usepackage{color}

\usepackage[colorlinks,linkcolor=blue,anchorcolor=black,citecolor=blue]{hyperref}
\begin{document}
	
	\title{5G Embraces Satellites for 6G Ubiquitous IoT: Basic Models for Integrated Satellite\\ Terrestrial Networks}
	
	\author{{Xinran Fang, 
			Wei Feng, \emph{Senior Member, IEEE},
			Te Wei, 
			Yunfei Chen, \emph{Senior Member, IEEE},\\
			Ning Ge, \emph{Member, IEEE},
			and Cheng-Xiang Wang, \emph{Fellow, IEEE}
		}
		\thanks{
			This work was supported in part by the National Key R\&D Program of China under Grant 2020YFA0711301, the National Natural Science Foundation of China (Grant No. 61771286, 61922049, 61941104, 61960206006), the Fundamental Research Funds for the Central Universities under Grant 2242019R30001, the EU H2020 RISE TESTBED2 project under Grant 872172, the Beijing
			National Research Center for Information Science and Technology project under Grant BNR2020RC01016, the Nantong Technology Program under Grant JC2019115, and the Beijing Innovation Center for Future Chip. \textit{(Corresponding author: Wei Feng.)}
			
			X. Fang, W. Feng, T. Wei, and N. Ge are with the Beijing National Research Center for Information Science and Technology, Department of Electronic Engineering, Tsinghua University, Beijing 100084, China. X. Fang, W. Feng, and T. Wei contributed equally to this work. T. Wei is also with the Department of WLAN Development, Huawei Beijing Research Center, Beijing 100085, China (e-mail:
			fxr20@mails.tsinghua.edu.cn;~fengwei@tsinghua.edu.cn;~wei-t14@tsinghua.org.cn;~gening@tsinghua.edu.cn).

			Y. Chen is with the School of Engineering, University of Warwick, Coventry CV4 7AL, United Kingdom (e-mail: Yunfei.Chen@warwick.ac.uk).
			
			C.-X. Wang is with the National Mobile Communications Research Laboratory, School of Information Science and Engineering, Southeast University, Nanjing, 210096, China, and also with the Purple Mountain Laboratories, Nanjing 211111, China (e-mail: chxwang@seu.edu.cn).
			
	}}
	
	\markboth{IEEE Internet of Things Journal, Vol. XX, No. XX, Month 2021}%
	{Shell \MakeLowercase{\textit{et al.}}: Bare Demo of IEEEtran.cls for IEEE Journals}

	\maketitle

	\begin{abstract}
		Terrestrial communication networks mainly focus on users in urban areas but have poor coverage performance in harsh environments, such as mountains, deserts, and oceans. Satellites can be exploited to extend the coverage of terrestrial fifth-generation (5G) networks. However, satellites are restricted by their high latency and relatively low data rate. Consequently, the integration of terrestrial and satellite components has been widely studied, to take advantage of both sides and enable seamless broadband coverage. Due to the significant differences between satellite communications (SatComs) and terrestrial communications (TerComs) in terms of channel fading, transmission delay, mobility, and coverage performance, the establishment of an efficient hybrid satellite-terrestrial network (HSTN) still faces many challenges. In general, it is difficult to decompose a HSTN into a sum of separate satellite and terrestrial links due to the complicated coupling relationships therein. To uncover the complete picture of HSTNs, we regard the HSTN as a combination of basic cooperative models that contain the main traits of satellite-terrestrial integration but are much simpler and thus more tractable than the large-scale heterogeneous HSTNs. In particular, we present three basic cooperative models, i.e., model \textit{X}, model \textit{L}, and model \textit{V}, and provide a survey of the state-of-the-art technologies for each of them. We discuss future research directions towards establishing a cell-free, hierarchical, decoupled HSTN. We also outline open issues to envision an agile, smart, and secure HSTN for the sixth-generation (6G) ubiquitous Internet of Things (IoT).
	
	\end{abstract}

	
	\IEEEpeerreviewmaketitle
	
	\section{Introduction}
	With the development of the fifth-generation (5G) communication networks, the world has witnessed a huge shift in the daily lives of people. People are not merely content to use the network to deliver messages but use it to interact with everything. Undoubtedly, the era of the Internet of Things (IoT) is around the corner. Numerous items, such as sensors, vehicles, tablets, and wearable devices, are joining the network, fostering a series of techniques and applications. For example, by leveraging the autonomous inspection of monitors, intelligent transportation \cite{n211,n89,n90}, coastal monitoring \cite{n91,Wei2021} and smart agriculture \cite{n92} are rapidly evolving. In addition, the agile measurement of sensors enables autonomous driving \cite{n93}, smart health care \cite{n94} and fast disaster recovery \cite{n95}. To accelerate the development of these applications, accompanying technologies such as machine learning \cite{n96}, unmanned aerial vehicles (UAVs) \cite{p4002,n97}, and blockchain \cite{n98} have been introduced to tackle communication, computation and security challenges. However, the items to be connected are widely distributed. For remote areas such as seas, mountains, and depopulated zones, traditional cellular base stations (BSs) are still difficult to deploy \cite{p4001}. In this sense, satellites could provide global coverage, and it is necessary to combine satellite communications (SatComs) and terrestrial communications (TerComs) to support the coming ubiquitous IoT.

    When discussing SatComs, there are several problems that need to be taken into account. First, the distance of a satellite link is much longer than that of a terrestrial link. Thus, the path loss of SatComs is usually very high, which requires ground terminals to be equipped with high-power transmitters and high-sensitivity receivers. As a result, it is difficult to keep terminals small. Second, the beam spots from adjacent satellites may overlap, resulting in severe inter-satellite interference. The cost of providing broadband communication services via satellites is very high. Thus, combining satellite and terrestrial networks to make use of the wide coverage of satellites and the high capacity of terrestrial networks in the sixth-generation (6G) era is of great interest \cite{p253}.
	
		\begin{table}[t]\centering
		\caption{Abbreviations.}\label{table-denotations}
		\begin{tabular}{|l|l|}
			\hline
			
			\textbf{Abbreviation} & \textbf{Full name} \\\hline
			3GPP& 3rd Generation Partnership Project \\\hline
			4G & fourth-generation \\\hline
			5G & fifth-generation \\\hline
			6G & sixth-generation \\\hline
			ADMM & alternating direction method of multipliers \\\hline
			AF	&	amplify-and-forward					\\\hline
			ASER&	average symbol error rate					\\\hline
			BCR	&	benefit-to-cost ratio					\\\hline
			BS	&	base station					\\\hline
			CCI	&	cochannel interference					\\\hline
			CNR & carrier-to-noise ratio \\\hline
			CSI & channel state information \\\hline
			DF	&	decode-and-forward					\\\hline
			EC	&	ergodic capacity					\\\hline
			eMBB & enhanced mobile broadband \\\hline
			FT & fixed terminal				\\\hline
			GEO	&	geostationary earth orbit	\\\hline
			GMT&	ground mobile terminal	\\\hline
			HAP & high-altitude platform	\\\hline
			HMT&	hybrid  mobile terminal					\\\hline
			HSTN	&hybrid satellite-terrestrial network							\\\hline
			ICN& information-centric networking\\\hline
			IoT& Internet of Things\\\hline
			IP & Internet Protocol \\\hline
			ITU& International Telecommunication Union \\\hline
			LEO	&	low earth orbit \\\hline
			MEC &  mobile edge computing 	\\\hline
			MEO	&	middle earth orbit 	\\\hline
			MGF & moment generating function \\\hline
			MIMO&	multiple input multiple output					\\\hline
			MISO&	multiple input single output					\\\hline
			MMSE & minimum mean squared error \\\hline
			mMTC & massive machine-type of communication 	\\\hline
			mmWave& millimeter wave \\\hline
			MPSK&	multiple phase shift keying					\\\hline
			MRC	&	maximum ratio combination					\\\hline
			NFV & network function virtualization \\\hline
			NOMA	&nonorthogonal multiple access \\\hline
			NR & new radio \\\hline
			NTN & non-terrestrial network 	\\\hline
			OP	&	outage probability					\\\hline
			PU & primary user \\\hline
			QoS & quality of service 	\\\hline
			RTT & round trip time \\\hline
			SaT5G & satellite and terrestrial network for 5G 	\\\hline
			SatCom & satellite communication \\\hline
			SC	&	selective combination					\\\hline
			SDN &  software defined networking \\\hline
			SER	&	symbol error rate					\\\hline
			SFT	&	satellite fixed terminal					\\\hline
			SIMO&	single input multiple output					\\\hline
			SINR & signal-to-interference-plus-noise ratio \\\hline
			SISO&	single input single output					\\\hline
			SMT&	satellite mobile terminal					\\\hline
			SNR & signal-to-noise ratio  \\\hline
			ST &satellite terminal \\\hline
			SU & second user  \\\hline
			TerCom & terrestrial communication \\\hline
			TBS & terrestrial base station \\\hline
			UAV & unmanned aerial vehicle \\\hline
		\end{tabular}
	\end{table}
	
    There have been a few key milestones in the conceptualization and development of hybrid satellite-terrestrial networks (HSTNs). The concept of HSTNs originated in 1964 \cite{5247781} and 1965 \cite{5247517,4206709}, where mutual interference between terrestrial BSs (TBSs) and fixed terminals (FTs) was studied. In 1983, Lee \textit{et al.} first introduced the concept of mobile satellite and terrestrial system symbiosis and discussed key issues \cite{1145889}. Later, in 1988, Richharia \textit{et al.} introduced the synergy of mobile satellites and terrestrial systems \cite{10414}. In 1992, Caini \textit{et al.} introduced a satellite-terrestrial system and evaluated the cochannel interference (CCI) \cite{166759}. The interference from terrestrial sources to satellite receivers was investigated in 1992 \cite{672027} and 1993 \cite{222474}.  In 1995, Ananasso \textit{et al.} considered the integration of SatComs and TerComs \cite{497076}. In 1996, Bond \textit{et al.} proposed the same idea as that in \cite{497076} from a business perspective \cite{864080}.
	
	Currently, with the development of 5G networks, the integration of satellites and 5G networks has attracted much attention from standardization organizations, companies and research institutes. Several organizations, such as the 3rd Generation Partnership Project (3GPP) and the International Telecommunication Union (ITU), have set up special working groups for the standardization of HSTNs. The ITU has proposed four application scenarios for satellite-5G integration and the key factors that must be considered to support these scenarios, such as intelligent routing and dynamic caching. The 3GPP has defined the deployment scenarios of non-terrestrial networks (NTNs) for 5G, including 8 enhanced mobile broadband (eMBB) scenarios and 2 massive machine-type communication (mMTC) scenarios~\cite{p275}. Some enterprises have also conducted research on satellite-terrestrial integration. In 2018, Satellite and Terrestrial Network for 5G (SaT5G) experimentally demonstrated the architecture of HSTNs, where a pre-5G test platform using software defined networking (SDN), network function virtualization (NFV) and mobile edge computing (MEC) technologies was integrated with geostationary earth orbit (GEO) satellites \cite{p248}. By February 2020, SaT5G had finished the 5G hybrid backhaul demonstration on the Zodiac Inflight Innovations testbed, which not only adopts network virtualization in both satellite and terrestrial components but also achieves integrated resource management and orchestration \cite{n2}. In September 2020,  European Space Agency announced the completion of the SatNex IV project and completed an early assessment of transforming promising terrestrial telecommunication technologies into space applications \cite{p276}. 

	\begin{figure*}[htbp]
	\centering
	\includegraphics[width=6.5in]{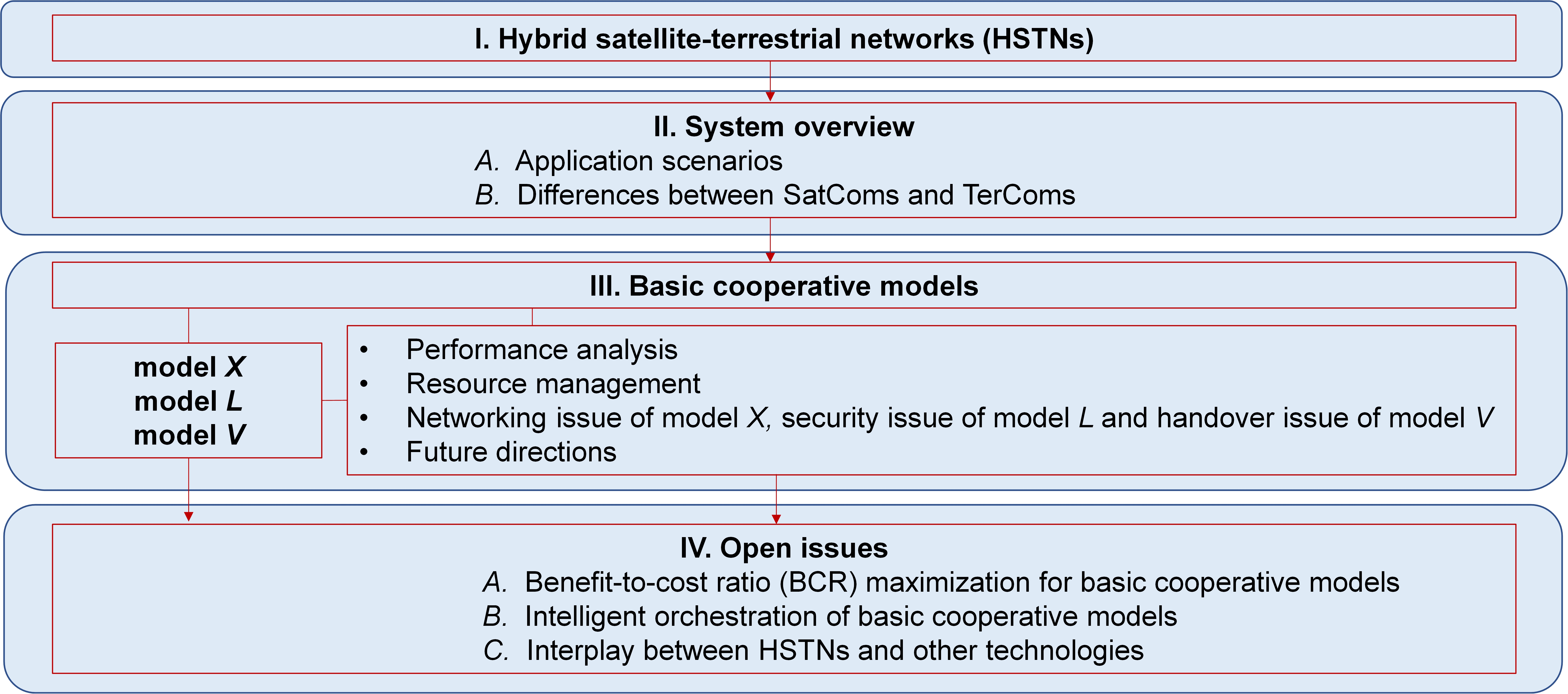}
	\caption{The structure of this paper.}
	\label{fig-content}
\end{figure*}
	
	To date, several survey papers have reviewed HSTNs from different perspectives. In particular, Burkhart \textit{et al.} investigated channel models and terrestrial interference for satellite television broadcasts \cite{7245586}. Focusing on the network and transportation layers, Taleb \textit{et al.} investigated the challenges, opportunities, and solutions of HSTNs \cite{5714025}. Niephaus \textit{et al.} conducted a survey on the quality of service (QoS) provision in HSTNs  \cite{7463014}. Liu \textit{et al.} reviewed network designs and optimization of HSTNs \cite{p269}. Wang \textit{et al.} provided a generic overview of the representative architectures, present research and evaluation works of different satellite-terrestrial networks \cite{n3}. These surveys have provided very useful discussions on the concepts, challenges, and key technologies of HSTNs. However, to the best of our knowledge, basic cooperative models for HSTNs have not been investigated. 
	
	Due to the significant differences between SatComs and TerComs in terms of channel fading, transmission delay, mobility, and coverage performance, 
	a large-scale HSTN cannot be simply decomposed into a sum of separate satellite and terrestrial links.
 	The gap between micro link analysis and macro network evaluation needs to be filled to uncover the complete picture of HSTNs.
 	Towards this end, we may consider the HSTN as a combination of basic cooperative models, that contain the main traits of satellite-terrestrial integration but are much simpler and, thus, more tractable than a large-scale heterogeneous HSTN.
	In this paper, we present a mesoscopic sight for HSTNs using three basic cooperative models, i.e., model $X$, model $L$, and model $V$, and provide a survey of the state-of-the-art technologies for each of them. We investigate some main problems and their solutions, from the perspectives of performance analysis, resource management, networking and security issues. On that basis, we point out future directions towards establishing a cell-free, hierarchical, decoupled HSTN. We also outline open issues to envision an agile, smart, and secure HSTN for 6G ubiquitous IoT. 

	The remainder of this paper is organized as follows. In Section II, an overview of HSTNs, including application scenarios and the main differences between SatComs and TerComs is presented. In Section III, we discuss three basic cooperative models to better understand HSTNs. For each model, the state-of-the-art technologies are further investigated, and future directions towards establishing a cell-free, hierarchical, decoupled HSTN are briefly discussed. Section IV outlines some open issues towards developing an agile, smart, and secure HSTN in the 6G era.  Finally, conclusions are given in Section V. The contents and architecture of this paper are shown in Fig. \ref{fig-content}. The abbreviations used in the paper are listed in Table \ref{table-denotations}.

	\section{System Overview}
	
	\begin{figure*}[htbp]
		\centering
		\includegraphics[width=6 in]{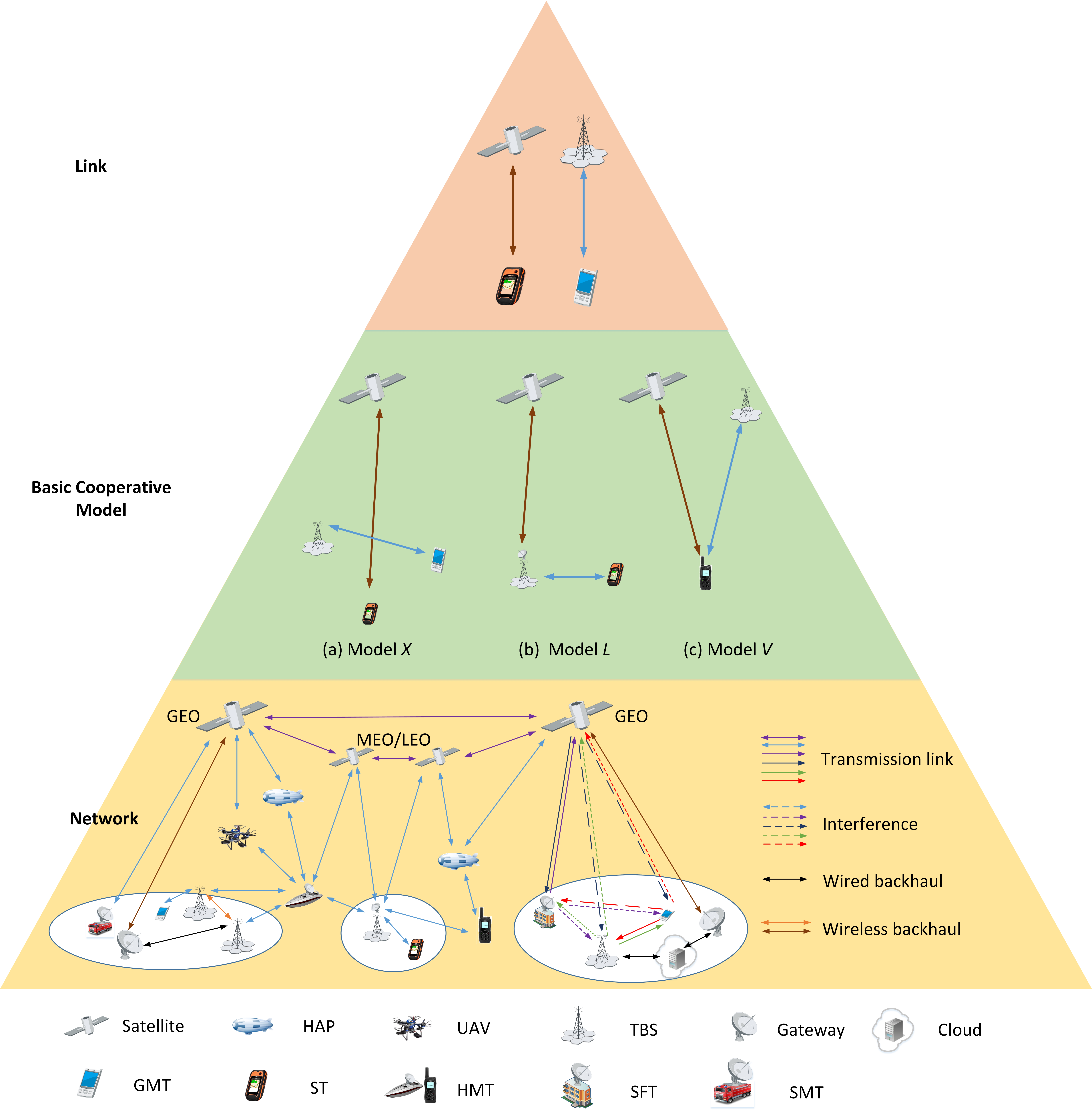}
		\caption{Illustration of a HSTN, which consists of satellite, aerial, and terrestrial domains. Focusing solely on wireless links, the HSTN can be considered as a combination of three basic cooperative models: model $X$, model $L$, and model $V$. These basic cooperative models provide a mesoscopic sight to fill the gap between micro link analysis (simple but difficult to characterize the coupling among diverse links) and macro network evaluation (including all interactions but complex) of HSTNs.}
		\label{fig-scenario}
	\end{figure*}
	
		\begin{table*}[t]
		\centering
		\caption{Comparison between SatComs and TerComs.}
		\begin{tabular}{|c|p{24em}|p{24em}|}
			\hline
			\multicolumn{1}{|c|}{\textbf{Characteristics}} & \textbf{SatComs} & \textbf{TerComs} \bigstrut\\
			\hline
			\multicolumn{1}{|c|}{\multirow{3}[18]{*}{Wireless channel}} & Higher propagation loss\newline{}Mainly affected by atmosphere and rain\bigstrut & Lower propagation loss\newline{}Mainly affected by blocks and scatters \bigstrut\\
			\cline{2-3}          & Mostly Rician channels (with a direct path) \bigstrut& Mostly multipath channels \newline{}(Rician channels in open areas) \bigstrut\\
			\cline{2-3}          & High Doppler frequency offset for MEO/LEO satellites\newline{}(e.g., approximately 35.4 kHz for Iridium) & Low Doppler frequency offset for low-speed GMTs \bigstrut\\
			\hline
			\multicolumn{1}{|c|}{\multirow{3}[6]{*}{One-way transmission delay}} & High\newline{}GEO satellites: approx. 270 ms\newline{}MEO satellites: approx. 130 ms (e.g., for O3b)\newline{}LEO satellites: less than 40 ms (e.g.,  10--30 ms for Globalstar) & Low\newline{}4G: less than 10 ms\newline{}5G: less than 1 ms \bigstrut\\
			\hline
			\multicolumn{1}{|c|}{\multirow{3}[2]{*}{Mobility}} & GEO satellites: static to earth\newline{}MEO/LEO satellites: fast (e.g., period less than 130 min for Globalstar) & Cellular communications: static \newline{} Device-to-device and vehicle-to-vehicle communications: up to hundreds of kilometers per second \bigstrut\\
			\hline
			\multicolumn{1}{|c|}{\multirow{3}[3]{*}{Coverage}} & Wide  & Limited \bigstrut[t]\\
			& Wide beam: over 100 km with a single beam for GEO satellites &  4G: 500--2000 m for a single cell in urban areas \\
			& Spot beam: depends on the beam width and altitude &  5G: 100--300 m  for a single cell in urban areas \bigstrut[b]\\
			\hline
		\end{tabular}%
		\label{table two}%
	\end{table*}%
	\subsection{Application Scenarios}
	As depicted at the bottom of Fig. \ref{fig-scenario},  the HSTN is an integration of satellite and terrestrial networks. TBSs, ground mobile terminals (GMTs) and backbone on the ground together make up the terrestrial network. The TBSs can access the cloud through wired backhaul. GEO satellites, medium/low earth orbit (MEO/LEO) satellites, satellite terminals (STs) including satellite mobile/fixed terminals (SMTs/SFTs), hybrid mobile terminals (HMTs)\footnote{In this paper, we refer to dual-mode mobile terminals, which can be used for both SatComs and TerComs, as HMTs.},  gateways, and high-altitude platforms (HAPs) make up the satellite network. In the HSTN, satellite networks and terrestrial networks are integrated together. Satellites can access the cloud from gateways \cite{p274}. In urban areas, cellular BSs and GMTs coexist with satellite receivers, and CCI is an important problem. In suburban areas, such as those near the sea, SatComs can be used jointly to provide seamless connections. HMTs can gain access from TBSs when they are within TBS coverage and can communicate via satellites when TBSs are not available. In remote regions, such as deserts, far sea, and rural areas, where cellular services are scarcely available, satellites can provide communication services, and TBSs usually work as relays to forward signals between satellites and STs \cite{p240}\cite{p241}. In summary, the incomplete coverage of terrestrial networks can be greatly strengthened in HSTNs through careful satellite constellation designs \cite{p252}. In addition, ultra-dense LEO networks can provide efficient data offloading \cite{p220}. Airships and airplanes can serve as high-altitude relays \cite{1391449}, and UAVs can provide complementary coverage \cite{n99}. Thus, the HSTN is composed of satellite, aerial, and terrestrial domains \cite{p217}\cite{p263}. 
	 
	Unlike a single satellite or terrestrial network, the HSTN is heterogeneous. The distinct characteristics of SatComs and TerComs impose great challenges when attempting to evaluate and establish an efficient HSTN. In the following subsection, the main differences between SatComs and TerComs are summarized.
	
	\subsection{Differences Between SatComs and TerComs}
	
	The SatCom radio propagation environment is quite different from that of the terrestrial case \cite{7245586,5264626}. For example, the transmission distance is longer (thus higher propagation loss and larger transmission delay), less scatterer occurs (thus usually with a direct path), and there are more attenuation effects due to rain and atmosphere conditions \cite{6493416,1137538,6772645}. Particularly, in \cite{6772645}, the influence on the coupling between satellite and terrestrial radios due to rain attenuation was analyzed. In \cite{6883690}, a unified multiple input multiple output (MIMO) channel model for mobile satellite systems with ancillary terrestrial components was presented. In \cite{p3008,p3007}, some approaches to predicting satellite channel statistics were proposed. The interference impact between satellite and terrestrial links was also discussed.

	In addition to channel models, there are also significant differences in transmission delay, mobility, and coverage performance of SatComs and TerComs. We summarized these differences of the fourth-generation (4G)/5G networks in Table \ref{table two}. Moreover, HSTNs need to serve a large number of users with various QoS requirements under limited spectrum and power resources. Resource reuse presents complex and varying interference under the influence of dynamic services, which directly restricts the system capacity and performance. Undoubtedly, the significant differences are the main factors restricting system performance. 
	
	Due to these differences, the HSTN cannot be simply decomposed into a sum of separate satellite and terrestrial links due to the complicated coupling relationships therein. However, it is also  impractical to treat the whole network as a unit due to the high complexity. To present a clear picture of the HSTN, we consider basic cooperative models of SatComs and TerComs, which contain the main traits of satellite-terrestrial integration but are much simpler and, thus, more tractable than the whole heterogeneous network.

	\section{Basic Cooperative Models}

	Although the network is complex, we find that HSTNs can be considered multifarious combinations of three basic cooperative models.  
	These basic cooperative models are expected to characterize the basic cooperation behavior between SatComs and TerComs with the fewest wireless links. As depicted in the middle layer of Fig. \ref{fig-scenario}, each model consists of only one satellite link and one terrestrial link, thus containing one satellite\footnote{Focusing on satellite-terrestrial integration, we draw only one satellite to represent a GEO/MEO/LEO satellite.}, one satellite user, one BS and one terrestrial user. For brevity, we name the basic cooperative models model $X$, model $L$, and model $V$.
	
	$\bullet$ Model $X$: A satellite and a TBS communicate with their users (ST/GMT) separately, sharing the same spectrum resources. The two lines in $X$ represent the satellite link and the terrestrial link.
	
	$\bullet$  Model $L$: A satellite communicates with its user via a relay, which serves as a combination of a BS and a ST. The three vertices in $L$ represent the satellite, the relay and the user.
	
	$\bullet$ Model $V$: A satellite cooperates with a TBS to serve one common user (HMT). The two lines in $V$ represent the satellite link and the terrestrial link, while the intersection denotes the HMT.
	
	The basic cooperative models, abstracted from various HSTNs, could fill the gap between micro link analysis and macro network evaluation, providing a mesoscopic sight to uncover the complete picture of HSTNs.
	In this way, an arbitrary HSTN can be considered a combination of these aforementioned cooperative models, and studies on each model will contribute to uncovering the complete picture of HSTNs.
	
	\begin{figure}[t]
		\centering
		\includegraphics[width=2.8in]{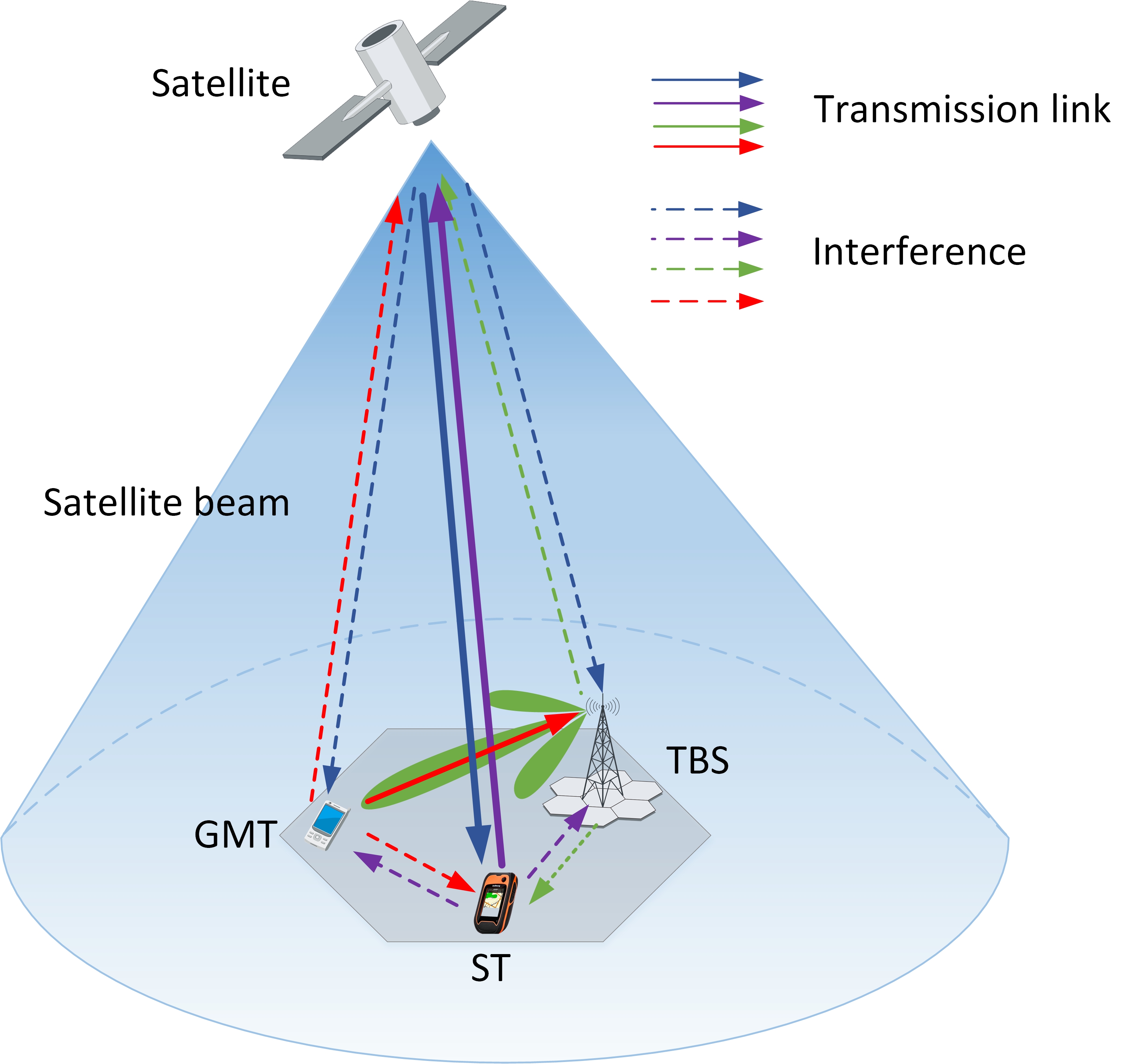}
		\caption{Illustration of the model $X$.}
		\label{fig3}
	\end{figure}

\begin{table*}[htbp]
	\centering
	\caption{Performance analysis for model $X$ and the major satellite-terrestrial differences considered.}\label{table-Performance-Model1}
	\begin{tabular}{|m{1.8cm}|m{2.6cm}|m{1.4cm}|m{2.2cm}|m{2.7cm}|m{4.6cm}|}
		\hline
		\textbf{Scenario} & \textbf{Channel difference (satellite/terrestrial)} & \textbf{Delay \newline{} difference} &\textbf{Interference type}&\textbf{Performance}& \textbf{Achievements and analytic tools} \\ \hline
		\multirow{11}{1.5cm}{Satellite to ST, SISO, downlink}
		& No (both free-space propagation loss)& & 	&Interference  measurement & Calculation formulas and measuring data on a vehicle \cite{640258}\\\cline{2-2}\cline{4-6}
		&\multirow{2}{2cm}{Yes (log-normal / cluster based scattering)} &\multirow{2}{2cm}{ }&\multirow{5}{2.2cm}{TBS interferes SFT}&EC of GMT& 	Meijer-G function based analytical expression,  with constrained interference for PU \cite{7510840} \\\cline{5-6}
		& & & & OP of GMT &Closed-form expression with constraints for PU, approximation of Gamma distribution \cite{7996553} \bigstrut \\\cline{2-2} \cline{4-6}
		&Yes (generalized-$K$ distribution/ gamma distribution)& &Satellite interferes PU, TBS interferes SU&Effective capacity of ST& Closed-form expression for the effective capacity of the secondary satellite network \cite{p515} \\\cline{2-2} \cline{4-6}
		
		&\multirow{3}{2cm}{Yes (Shadowed-Rician / Nakagami-$m$)}  &&\multirow{3}{2.2cm}{Satellite interferes cellular users, TBS interferes SFT} &EC of NOMA users&Analytical expression of EC based on the MGF \cite{p514}\\ \cline{5-6}
		
		&   &{\color{white} multiple TBSs} &	&OP of GMT & Closed-form expression \cite{7898463}\\\cline{2-2}\cline{4-6}

		& {Yes (Shadowed-Rician /  Suzuki)} &{\color{white} multiple TBSs} &TBS interferes SFT, satellite interferes GMT & OP  and EC of SFT and GMT, diversity/coding gain &  Unified closed-form analysis, the  diversity/coding gain relationship, the fading parameter, the shadowing parameter \cite{7248442} \bigstrut
		\\\cline{1-2}\cline{4-6}
		
		\multirow{ 1}{1.8cm}{SFT to satellite, SIMO, uplink}
		& \multirow{3}{2cm}{ No (both Rician fading channels) }&\multirow{2}{2cm}{} &\multirow{2}{2.2cm}{TBS interferes satellite}&\multirow{3}{*}{Capacity of the satellite}	& Closed-form approximation of the upper bound capacity and the capacity with linear MMSE  \cite{7556323} \\\cline{6-6}
		& & & & & Approximation of optimal joint decoding capacity and MMSE capacity with Haar measuring\cite{7878033}		\\\cline{1-2} \cline{4-6}
		SFT to satellite, SISO, uplink & No (both rain-fading with log-normal distributions)&\multirow{2}{*}{} &SFT  interferes terrestrial receiver &Capacity of the satellite	& Interference analysis of the terrestrial link and capacity analysis of the satellite link  \cite{n75} \\\cline{6-6}
		\hline
	\end{tabular}		
\end{table*}

\begin{table*}[t]\centering
	\caption{Resource management for model $X$ and the major satellite-terrestrial differences considered.}
	\label{table-ResourceManagementx}
	\begin{tabular}{|m{2cm}|m{2.5cm}|m{8cm}|m{1cm}|m{1cm}|}			
		\cline{1-5}
		\textbf{Goal}  & \textbf{Schemes} & \textbf{Achievements and analytic tools} & \textbf{Channel\newline{}difference} & \textbf{Delay\newline{}difference}\\
		\cline{1-5}
		\multirow{8}{*}{Capacity}
		&\multirow{17}{*}{Power allocation}
		&		Effective capacity under QoS requirements, with both perfect and imperfect CSI considered\cite{6517359} & Yes & Yes  \\ \cline{3-5}
		&&Power control of STs to reduce the interference affecting terrestrial links\cite{7391092} & Yes & No \\\cline{3-5}
		&      & Spectrum reuse between satellite beams and terrestrial cells, power allocation to mitigate interference according to the traffic demand\cite{7373246} & Yes & No \\\cline{3-5}
		&      &	Delay-limited capacity under average and peak power constraints\cite{7882651}&  Yes & Yes \\\cline{3-5}
		&      &	Power allocation of both the satellite and TBSs in a distributed way based on game theory \cite{n41} & Yes & No \\ \cline{3-5}
		&      &	Power control schemes from long-term and short-term perspectives to tackle spectrum sharing problems \cite{n71} & Yes & No \\\cline{1-1} \cline{3-5}
		\multirow{6}{*}{Energy}
		&&Power allocation scheme of the satellite network under  interference  and delay constraints, closed-form OP expressions based on the obtained optimal transmit power \cite{p3001} & Yes & Yes  \\\cline{3-5}
		&  &Power control of the satellite network with the interference constraint of PUs and the minimal rate requirement of SUs based on outdated CSI \cite{n4} & Yes & No  \\\cline{1-5}
		\multirow{13}{*}{Capacity}
		&\multirow{15}{2.5cm}{Spectrum sharing  and  carrier allocation}
		&      	Improving the spectrum efficiency in the  S band by beamforming, spectrum coordination between satellite and terrestrial components\cite{5586925} & Yes & No \\\cline{3-5}
		&  &    Database approach \cite{7248431}& Yes & No \\\cline{3-5}
		
		&      &	Spectrum sharing using large-scale CSI \cite{7928993} & Yes & No \\\cline{3-5}
		&       &Joint beamforming and carrier allocation for SatComs  \cite{7336495} & Yes & No \\\cline{3-5}
		&      &	Sequential carrier allocation between satellite and terrestrial systems\cite{7962743} & Yes & No \\ 
		\cline{3-5}	
		&      &Carrier and power allocation with imperfect CSI, a dual decomposition method \cite{7986309} & Yes & No \\ \cline{3-5}
		&      &Joint power and subchannel allocation of the satellite uplink with the OP constraint of terrestrial users \cite{n42} & Yes & No \\ \cline{1-1}  \cline{3-5}	
		Delay&      &Power and bandwidth allocation based on nonideal sensing  in a distributed manner \cite{p219} & Yes & No \\ \cline{1-5} 
		\multirow{6}{1.5cm}{Interference Mitigation}&\multirow{18}{*}{Beamforming}&Both adaptive and non-adaptive beamforming schemes, a robust gradient based switching mechanism \cite{6294500}&-&No \\\cline{3-5}
		&&Simulation of adaptive beamforming \cite{7058570} &-&No\\\cline{3-5}
		&&Joint carrier allocation and beamforming scheme based on the radio environment map \cite{p506}&No&No\\\cline{3-5}\cline{1-1}
		
		\multirow{2}{1.5cm}{Fairness}&&Optimal beamforming weight vectors and power allocation using the uplink-downlink duality theory \cite{p232}&Yes&No\\\cline{3-5}
		&&Beamforming with interference and power transfer requirements \cite {p268}&-&No\\\cline{3-5}\cline{1-1}
		
		Energy&&Beamforming design, a UAV eavesdropper \cite{n58}&No&No \\\cline{1-1} \cline{3-5}
		\multirow{5}{1.5cm}{Capacity}&&Single user and multi-user analog-digital beamforming schemes as well as hybrid carrier allocation \cite{7980751} &No&No\\\cline{3-5}
		&&User pairing scheme for NOMA users, joint power allocation and beamforming optimization \cite{n43}  &Yes&No\\\cline{3-5}
		&&Beamforming design of  the uniform planar array for TBS, OP and EC analysis\cite{n64}
		&Yes&No\\\cline{3-5} 
		&&Beamforming design of the satellite with a nonlinear power amplifier based on large-scale CSI\cite{n5}	&-&No\\   \cline{1-1}  \cline{1-5}
	\end{tabular}
\end{table*}

	\subsection{Model $X$}
	As shown in Fig. \ref{fig3}, in model $X$, a satellite and a TBS share spectrum resources to communicate with the ST and the GMT, respectively. One of the key technical challenges of this model lies in the mutual interference between the satellite link and the terrestrial link \cite{p501}\cite{p502}. In particular, the interference from terrestrial users to satellites was studied by infield measurements and simulations in \cite{7417497}.
	Different from the CCI between two links within a pure satellite or terrestrial network, the interference patterns in model $X$ are diverse and complicated due to the aforementioned satellite-terrestrial differences.

	The main interference in urban and rural areas differs greatly due to the different coverage of a single beam/cell of SatComs and TerComs \cite{p504}. In urban areas, STs mainly suffer the interference from adjacent TBSs, which is more impactful than the interference from GMTs. For satellites, the interference from  GMTs is almost negligible compared to that from TBSs. For TBSs and GMTs, the interference from neighboring STs is usually much stronger than that from satellites, as satellite signals generally endure more severe attenuation. The gateways may also suffer interference from TBSs when they are close to each other. In this case, the deployment of TBSs and gateways should be jointly optimized to mitigate harmful interference \cite{7343438}. 
	In rural areas, the interference between STs and TBSs/GMTs can be ignored because they are usually separated from each other by great distances. 
	For satellites, the interference from TBSs is dominant, while that from GMTs can be ignored \cite{6166516}. 
	For TBSs/GMTs, the interference from satellites can be neglected, as it is much smaller than the desired signal.
	
	Next, we discuss existing studies on model $X$ in terms of system performance, resource management and the networking issue.
	
	\subsubsection{Performance Analysis}
	Some papers have analyzed the performance of model $X$ in terms of capacity, ergodic capacity (EC), and outage probability (OP). We summarize them in Table \ref{table-Performance-Model1}. In \cite{640258}, both theoretical and experimental methods were applied to measure the interference between fixed satellite services and terrestrial radio-relay services. In \cite{7556323} and \cite{7878033}, the capacity of satellite links with Rician fading was presented. In \cite{7556323}, an upper bound capacity of single input multiple output (SIMO) uplinks from SFTs to the satellite was given. In\cite{7878033}, both optimal joint decoding capacity and linear minimum mean square error (MMSE) capacity were derived considering satellite inter-beam interference and TBS interference.

	To some extent, model $X$ can be recast as a special case of cognitive radio. In \cite{7510840,7996553}, a cognitive network where the ST acted as the primary user (PU) and the GMT acted as the second user (SU) was considered. The EC of GMTs was derived in \cite{7510840} and a closed-form OP expression for a single input and single output (SISO) downlink was derived in \cite{7996553}. In contrast, the case in which a terrestrial network acted as the primary system, sharing the spectrum with the satellite was studied in \cite{n51}. The authors investigated the OP and EC of satellite uplinks under the consideration of imperfect channel state information (CSI). 
	In \cite{p514}, Yan \textit{et al.} investigated the EC performance of terrestrial users under the non-orthogonal multiple access (NOMA) transmission scheme and the superiority of the NOMA-assisted HSTN was proofed.  In \cite{p515}, Ruan \textit{et al.} considered the interference from the satellite to the GMT and the interference from TBSs to the ST and derived a closed-form expression of effective capacity based on the moment generating function (MGF).

	To directly characterize the interference in model $X$, the performances of both the ST and the GMT with interference were studied in \cite{7248442}, where the relationships among diversity gain, fading and shadowing parameters were presented. For an extension of model $X$, in \cite{7898463}, the interference from both the satellite and the adjacent TBSs to cellular users was considered, and a closed-form expression of OP was derived. In light of the standard recommendations of the ITU, the interference levels of terrestrial fixed services and the capacity of fixed satellite services were analyzed, offering a useful guideline for efficient designs of satellite-terrestrial coexistence \cite{n75}.
	
	Note that in the above studies, the Shadowed-Rician fading channel was the most widely used channel model between the satellite and the ground station \cite{7898463, p515, 7248442}. However, comprehensive performance evaluations under the general Nakagami-$m$ channel model remain open. In addition, it is difficult to acquire perfect CSI in practice, which causes great difficulty in the implementation of accurate interference evaluations. Further studies with more practical analyses need to be explored.

	\subsubsection{Resource Management}
	To enhance the spectrum efficiency, the spectrum is usually shared in model $X$. In this case, methods used to manage radio resources, including the spectrum, power and beams, greatly influence the system performance. We summarize the main literature on resource management for model $X$ in Table \ref{table-ResourceManagementx}.

	\paragraph{Power allocation}
	The existing studies on power allocation still focus mainly on capacity promotion.
	Vassaki \textit{et al.} proposed a power control algorithm under the QoS requirement \cite{6517359}.
	Lagunas \textit{et al.} studied a power control scheme where the satellite uplink and terrestrial downlink coexist in the Ka band \cite{7391092}.
	Park \textit{et al.} proposed a power allocation scheme to mitigate the inter-component interference between satellite beams and terrestrial cells \cite{7373246}.
     Gao \textit{et al.} proposed an alternating direction multiplier method (ADMM)-based power control scheme to optimize the uplink throughput \cite{p237}.

    In addition to merely optimizing the capacity, some works have also jointly considered other goals to systematically design the power allocation strategy.  
	Shi \textit{et al.} investigated two power control schemes to optimize the delay-limited capacity and the OP for real-time applications \cite{7882651}.
	To reduce system costs of centralized processing, Chen \textit{et al.} proposed a distributed power allocation scheme based on game theory \cite{n41}. 
	By taking the mobility of LEO satellites into account, Hu \textit{et al.} investigated a power control scheme to simultaneously maximize the capacity and minimize the OP  \cite{n71}.
    Wang \textit{et al.} investigated a joint power allocation and channel access scheme to maximize the rate of terrestrial users \cite{n78}. 
	Moreover, Hua \textit{et al.} considered a UAV-assisted HSTN and optimized the transmit power of the TBS/UAV \cite{p208}.
	
	Energy efficiency is also very important for HSTNs because the payload of a satellite is always limited.
	In \cite{p511}, Ruan \textit{et al.} proposed a power allocation scheme for a cognitive satellite-vehicular network to provide a tradeoff between energy efficiency and spectrum efficiency.
	In \cite{p3001}, an energy efficient power allocation strategy was proposed for cognitive HSTNs under both delay and interference constraints.
	Based on outdated CSI, the energy efficient power allocation scheme was further investigated in \cite{n4}, where the interference constraint of the terrestrial components and the minimal rate requirement of the satellite networks were taken into account.

	\paragraph{Spectrum sharing and carrier allocation}
	To alleviate the spectrum scarcity problem, Deslandes \textit{et al.} studied a spectrum sharing strategy, and the mutual interference between the satellite link and terrestrial link was considered \cite{5586925}. Tang \textit{et al.} applied a database approach for spectrum sharing in the Ka band \cite{7248431}.
	Feng \textit{et al.} proposed a large-scale-CSI-based spectrum sharing strategy for HSTNs \cite{7928993}.
	Zhu \textit{et al.} designed a resource allocation algorithm to reduce the interference based on imperfect CSI \cite{7986309}.
	By applying the exclusive zone for interference mitigation, Jia \textit{et al.} proposed a cognitive spectrum sharing and frequency reuse scheme to improve the energy efficiency and ensure inter-cell fairness \cite{p260}.
    Based on non-ideal spectrum sensing, Wang \textit{et al.} provided a distributed resource allocation algorithm  \cite{p219}. 
	Considering both downlink and uplink transmissions, Lagunas \textit{et al.} presented a joint beamforming and carrier allocation scheme for the satellite downlink,  and a joint power, carrier and bandwidth allocation scheme for the satellite uplink \cite{7336495}. Further in \cite{7962743}, Lagunas \textit{et al.} considered a wireless backhaul scenario and proposed a carrier allocation scheme for the enhancement of the overall spectral efficiency.
	Recently, Chen \textit{et al.} proposed a joint power and bandwidth allocation scheme to achieve a tradeoff between fairness and efficiency  \cite{n42}.

	\paragraph{Beamforming}
	The beamforming scheme could mitigate severe interference and combat  high path loss. Combined with the technique of millimeter wave (mmWave) communications, beamforming could further improve the spectrum efficiency \cite{n83}. 
	For uplink transmissions \cite{6166516,5286394,6294500,7058570}, the iterative turbo beamforming scheme\cite{6166516}, adaptive beamforming scheme \cite{5286394}, and semi-adaptive beamforming scheme \cite{6294500} were investigated for interference mitigation. For the case of downlink transmissions, 
	a multiple input single output (MISO) scenario was considered in \cite{6636830} with the goal of signal-to-interference-plus-noise (SINR) maximization. Joint optimization studies were conducted in \cite{p506,7980751,p267,p232}. Combined with carrier allocation \cite{p506,7980751}, BS deployment \cite{p267} and power allocation \cite{p232}, the beamforming schemes were jointly designed to tackle CCI.

    Moreover, some novel techniques and practical assumptions were considered for advanced beamforming. In \cite{p268,p272}, the beamforming schemes were applied to enhance secure transmissions, and the minimal achievable secrecy rate as well as total transmit power was optimized, respectively. 
	Furthermore, Lin \textit{et al.} investigated a joint beamforming scheme of satellite and terrestrial components where ground users were grouped into clusters based on the NOMA technique \cite{n43}.
	To explore the mmWave band in HSTNs, Zhang \textit{et al.} optimized the downlink beamforming of TBSs with the interference probability constraint of satellite users \cite{n64}. 
	Ruan \textit{et al.} considered a scenario where the UAV acts as a malicious eavesdropper \cite{n58}. 
	More recently, Liu \textit{et al.} investigated downlink beamforming with a nonlinear power amplifier \cite{n5}. By utilizing environmental and location information,  Wei \textit{et al.} optimized the precoding matrix of maritime users to maximize the ergodic sum capacity \cite{n100}.

\subsubsection{Networking Issue}
If  we extend the satellite and terrestrial links in model $X$ to a satellite sub-network and a terrestrial sub-network, respectively, we may face networking issues of integrating two very different networks. 
To address this point, some promising architectures have been proposed in the literature. In particular, Guidotti \textit{et al.} presented an architecture of HSTNs and analyzed the main technical challenges caused by satellite channel impairments, such as large path losses, delays, and Doppler frequency shifts \cite{p204}.

Chien \textit{et al.} introduced the architecture and challenges of HSTNs for IoT applications, where 5G, Wi-Fi, Bluetooth, LoRa and other transmission technologies were jointly exploited \cite{p211}.
Huang \textit{et al.} presented the evolution of wireless communications towards 6G and provided an entire architecture of the future green network, including ground, aerial, space and undersea components \cite{n88}. In addition, Charbit \textit{et al.} provided a system design for the narrowband IoT air interface of HSTNs, aiming to construct a backward compatible solution for future various IoT applications \cite{n86}. Liu \textit{et al.}  designed a new task-oriented intelligent networking architecture including space, air, ground, aqua components \cite{n85}. By introducing edge cloud computing, intelligent methods and information center networks, the network has the potential to tackle the challenges of intelligent networking, heterogeneous network interactions, intermittent network interruptions, long latency, and load unbalance. However, the authors only  provided a macro outlook of this heterogeneous architecture and many detailed schemes have not yet been discussed.

{To facilitate the deployment of new techniques and keep pace with the fast evolution of communication systems, SDN and NFV techniques can be utilized for satellite-terrestrial integration. Bertaux \textit{et al.}  pointed out that  programmability, openness, and virtualization are new trends for HSTNs \cite{n23}. With this consensus, Boero \textit{et al.} estimated the end-to-end delay \cite{p254},
Lin \textit{et al.} studied virtual spectrum allocation techniques \cite{p215} and Niephaus \textit{et al.} investigated dynamic traffic offloading \cite{p249} for SDN-based HSTNs. Zhang \textit{et al.} implemented MEC techniques in HSTNs and discussed a task offloading model to improve the QoS of mobile users \cite{p209}. Bi \textit{et al.} proposed some possible ways to improve the QoS using edge computing \cite{p216}. Feng \textit{et al.} creatively devised a flexible architecture named HetNet, which synthesizes the locator/ID split and information-centric networking (ICN) to accelerate the convergence of satellite and terrestrial networks \cite{n102}. To satisfy the
growing diversity of users' needs, Feng \textit{et al.} further  introduced service function chains into HSTNs and designed an efficient service function chain mapping approach \cite{n103}. By introducing SDN techniques into LEO satellites, Papa \textit{et al.} investigated controller placement and satellite-to-controller assignment to overcome the impact of topology changes and traffic variations  \cite{n82}. 
Based on virtualization technology,
Chen \textit{et al.} introduced the resource cube to depict the minimum unit of resources and designed a service-matching scheme to minimize the total system delay \cite{n80}. Wang \textit{et al.} assumed that virtual resources could be embedded into any physical nodes of the HSTN and proposed a traffic scheduling scheme to effectively allocate resources\cite{n84}. Recently, Feng \textit{et al.} further considered the elastic resilience of SDN-enabled HSTNs \cite{n104}. The controller reachability issue, the failure detection and recovery problems have been discussed in depth.}

	\begin{figure}[t]
	\centering
	\includegraphics[width=2.8in]{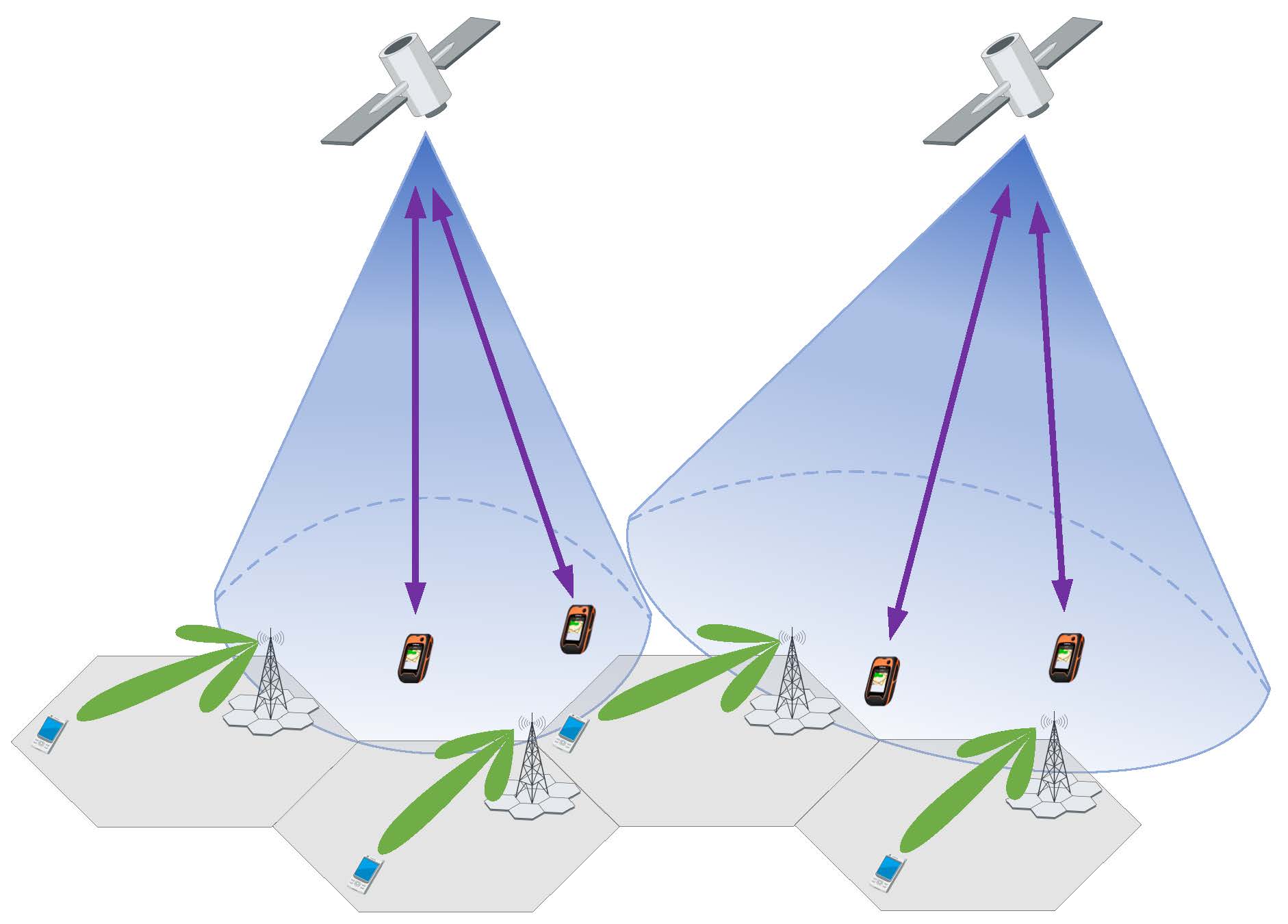}
	\caption{Illustration of a three-dimensional cell-free HSTN, derived from the extension of model $X$.}
	\label{fig_cellfree}
\end{figure}

	\begin{figure}[t]
	\centering
	\includegraphics[width=2.8in]{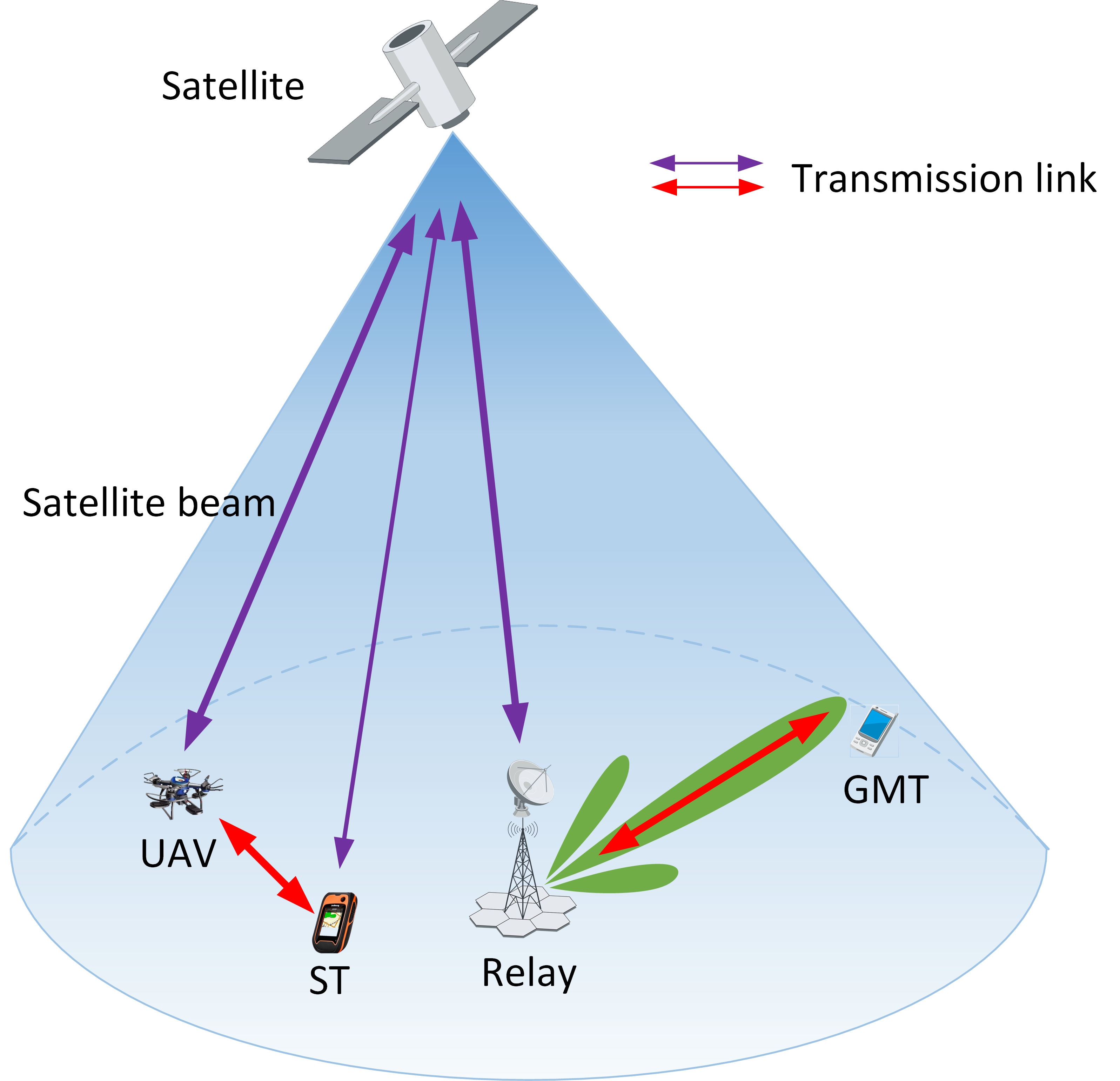}
	\caption{Illustration of the model $L$.}
	\label{fig4}
\end{figure}

	\subsubsection{Future Directions}
	Most existing works on model $X$ have focused on the satellite-terrestrial differences in wireless channels and coverage performance, while the delay difference has been mostly ignored, which should attract more research attention. Taking all the differences into account, we may further uncover the interference mechanism in model $X$. 
	On that basis, systematic system evaluation and holographic resource management can be envisioned. We also want to emphasize that
	practical constraints, including limited system cost, non-linear hardware, and imperfect prior knowledge, should be carefully handled in future studies.
	
	As shown in Fig. \ref{fig_cellfree}, the extension of model $X$ with multiple satellite/terrestrial links no longer follows the cellular architecture, which is regular in general. Thus, the cellular network could be decoupled into geographically separated cells. However, for HSTNs, satellite-terrestrial integration leads to a three-dimensional cell-free network architecture. Due to the complicated coupling therein, the cell-free architecture is undecomposable and difficult to analyze. One possible solution is to dynamically decouple the network on radio maps instead of on the real geographical map. This radio map may characterize the large-scale interference relationship. However, the basic theory and methods still remain unknown.

	\begin{table*}[htbp]\centering
	\caption{System performance for model $L$ and the major satellite-terrestrial differences considered.}\label{table-Performance-Model2}
	\begin{tabular}{|m{1cm}|m{1cm}|m{1.5cm}|m{1.7cm}|m{1.9cm}|m{1.5cm}|m{5.5cm}|}
		\hline
		\textbf{Relay mode} & \textbf{Relay number} & \textbf{Performance}& \textbf{Satellite link}& \textbf{Terrestrial link}& \textbf{Direct link}&\textbf{Achievements and analytic tools}  \\\hline
		\multirow{35}{*}{AF}
		&\multirow{20}{*}{Single}
		&\multirow{5}{*}{ASER}&\multirow{20}{1.3cm}{Shadowed-Rician}&Free-space optical link, \newline{} Gamma-Gamma fading &No&ASER for MPSK, analytical diversity order \cite{6605500}  \\\cline{5-7}
		
		&	&	&& \multirow{12}{*}{Nakagami-$m$}&Yes & ASER for MPSK, analytical diversity order\cite{6589295} \\\cline{6-7}
		
		&	& & &         &Yes& Both single relay and multiple relay networks are considered \cite{7308010} \\\cline{3-3}\cline{6-7}
		
		&	&ASER, average capacity \bigstrut&        &          &No& Channel estimation and detection designs, MGF with imperfect CSI \cite{7247422} \\\cline{3-3}\cline{6-7}
		
		&	&\multirow{3}{*}{OP} & &   &Yes& Closed-form and asymptotic expressions of OP of the NOMA-aided HSTN \cite{n39} \\ \cline{6-7}
		
		&	& &     &          &No &OP evaluation under opportunistic user scheduling \cite{7561027}  \\  \cline{3-3}\cline{6-7}	
		
		&	&OP, EC  &    &             &No & Approximated expression, lower and upper bound of EC, asymptotic  OP \cite{7185366}  \\\cline{3-3} \cline{5-7}
		
		&	&ASER &      &        &No&Beamforming scheme of the multi-antenna relay, analytical diversity order \cite{6784552}  \\\cline{3-3}\cline{6-7}
		
		&	&  OP, SER\bigstrut& &\multirow{10}{*}{Rayleigh}       &Yes & Alamouti code\cite{7504407}  \\\cline{3-3}\cline{6-7}
		
		&	&OP, ASER, EC &&  &No &S-R: MISO, S-D: MIMO, R-D: SIMO, closed-form expression of EC \cite{7230282} \\\cline{3-3}\cline{6-7}
		
		&	& OP, throughput &     &         &No &OP and throughput analysis under the effect of hardware impairments\cite{7918502}
		\\\cline{3-4} \cline{6-7}
		
		&	&SER \bigstrut&\multirow{4}{*}{Rician} &        &Yes&Closed-form expression of SER \cite{5586888}\\\cline{2-3} \cline{6-7}

		&\multirow{15}{*}{Multiple}
		&OP\bigstrut& &         &Yes&Closed-form expressions of the OP under non-identical relay channels\cite{5947821}  \\\cline{3-4}\cline{6-7}
		
		&	&\multirow{2}{*}{ASER}&Shadowed-Rician       &           &No \bigstrut &Analytical expression of the MGF with CCI, MRC at destination \cite{7094777} \\\cline{3-4}\cline{5-7}

		&	&OP, SER, achievable rate &$\upkappa-\upmu$ shadowed fading  &\multirow{15}{*}{Nakagami-$m$}        &No&Performance analysis of single and multiple relays under two relay selection schemes \cite{n67}\\\cline{3-4} \cline{6-7}

		&    &\multirow{6}{*}{OP} &\multirow{16}{1.3cm}{Shadowed-Rician}         &        &Yes/No& Two underlying selection policies to minimize the OP \cite{7981353} \\ \cline{6-7}
		
		&	&  &      &           &No & S-R: MISO, S-D: MISO, R-D: SISO, closed-form expression of OP, user and relay selection \cite{7332924} \\\cline{6-7}

		&	&  &      &          &No& OP, diversity analysis of two interference scenes \cite{n27} \\\cline{3-3}\cline{6-7}
		
		&	&OP, EC, SER &       &        &Yes &MIMO-enabled relay, OP and probability of error analysis of three transmission cases \cite{n70}\\\cline{1-3}\cline{6-7}

		\multirow{14}{*}{DF}
		&\multirow{6}{*}{Single}
		& \multirow{6}{*}{OP}&       &           &No &OP of satellite-relay-destination with beamforming performed in the relay\cite{7098351} \\ \cline{6-7}
		
		&	&  &        &          &No &OP analysis under imperfect CSI, power allocation to ensure user fairness 	\cite{n79} \\\cline{5-7}
		
		&	& &      & \multirow{8}{*}{Rayleigh}          &No &OP and energy efficiency analysis using a NOMA-enabled relay 	\cite{n69} \\\cline{2-3}\cline{6-7}
		
		&\multirow{8}{*}{Multiple}
		&OP, EC &       &   &Yes & Relay selection scheme, analytical diversity order  \cite{n38} \\\cline{3-4}\cline{6-7}
		
		&&SER  &\multirow{2}{1.5cm}{Land mobile satellite fading}     &    &Yes &Selective decode-and-forward transmission,  closed-form expression of  symbol error probability using MGF \cite{6449258}  \\\cline{3-3}\cline{6-7}
		
		&  &OP      &  &  &Yes &Best relay selection and analytical expression of OP using MRC and MGF \cite{6266136}, closed-form expression of OP\cite{6655280} \\\cline{3-7}
		
		&  &EC \bigstrut & Rician & Nakagami-$m$  &Yes & Closed-form expression of  EC \cite{7343591}  \\\cline{1-7}
		
	\end{tabular}
	
\end{table*}

	\begin{table*}[t]\centering
	\caption{Resource management for model $L$ and the major satellite-terrestrial differences considered.}
	\label{table-ResourceManagementL}
	\begin{tabular}{|m{2cm}|m{2.6cm}|m{8cm}|m{1cm}|m{1cm}|}			
		\cline{1-5}
		\textbf{Goal}  & \textbf{Schemes} & \textbf{Achievements and analytic tools} & \textbf{Channel\newline{}difference} & \textbf{Delay\newline{}difference}\\
		\cline{1-5}
		
		\multirow{5}{*}{Energy} &Power allocation and relay selection& Multi-relay, mixed binary and fractional optimization problem, binary relaxation  and dual decomposition \cite{7794885} & Yes &No\\
		\cline{2-5}
		&Power allocation and transmit model selection&SER performance of two transmission modes,  adaptive transmission scheme to maximize the energy efficiency \cite{p513} & No& No\\\cline{2-5}
		&\multirow{2}{2.5cm}{Power and subchannel allocation}& Power allocation for the satellite and satellite-terrestrial terminal, subchannel allocation for the ground downlink offloading, a binary-search-aided algorithm\cite{n40} & Yes & No\\
		\cline{1-1} \cline{3-5}
		Capacity&	&Maximize
		the sum throughput  under delay	constraints of delay-sensitive services \cite{n6} & Yes & Yes \\\cline{2-5} 
		&\multirow{3}{*}{Beamforming}&Beamforming design with multiple relays, multiple GMTs and multiple eavesdroppers \cite{7506598} & Yes & No \\\cline{1-1}\cline{3-5}
		Energy&&Beamforming design of the  UAV relay with angular-information-based CSI \cite{n36}& Yes & No\\\cline{1-5}	
	\end{tabular}		
\end{table*}

	\subsection{Model $L$}
		As shown in Fig. \ref{fig4}, model $L$ typically consists of one satellite, one relay and one destination user. In addition to the TBS, aerial facilities, such as UAVs, can play the same role as relays. The destination user can be either a GMT, which cannot directly receive the signal from the satellite, or a ST, which has a direct transmission link from the satellite. Compared with terrestrial/aerial links, direct satellite links are usually weaker due to limited size of mobile terminals.
		In this model, the relay can enhance the satellite links. This is especially important for users in remote rural, desert and sea areas. These areas are out of the coverage of terrestrial 4G/5G networks and mainly rely on satellites for communications. For users that can only achieve narrowband satellite services or are unable to directly access the satellite, the relay could provide broadband connections using model $L$. As an extension of basic model $L$, when multiple relays are considered, efficient relay selection could help improve the on-demand coverage extension. Similar to model $X$, the differences between SatComs and TerComs in wireless channels and beam/cell coverage impact the performance of model $L$. The delay difference becomes a critical factor concerning the two-phase transmission in model $L$. Next, we discuss the related works on model $L$ from the perspectives of system performance, resource management and the security issue.

	\subsubsection{Performance Analysis}
	Some papers have analyzed the performance of HSTNs under model $L$ in terms of symbol error rate (SER), average symbol error rate (ASER), capacity, EC and OP. We summarize the related studies in Table \ref{table-Performance-Model2}. According to relaying modes, the existing works can be classified into two categories, namely, the Amplify-and-Forward (AF) relaying type and the Decode-and-Forward (DF) relaying type. In the AF mode, the relay amplifies the received signal from the satellite and then forwards it to the destination. 
	In the DF mode, the relay decodes the received signal and forwards the decoded information to the destination. Compared with conventional AF/DF relay systems with homogeneous links, the  enormous differences between SatComs and TerComs pose new challenges for model $L$.

	Most of the existing literature focuses on the performance of model $L$ achieved by the one-way relay, under which the satellite transmits to the relay in the first phase, and the relay forwards the received signal to the destination in the second phase.
	In \cite{5586888}, the SER was derived with AF relays over non-identical fading channels.
	In\cite{7504407}, the authors analyzed the OP and SER of the Alamouti HSTN.
	In \cite{6589295} and \cite{6605500}, the multiple phase shift keying (MPSK) ASER was derived for different terrestrial channels. 
	Extending to the multi-antenna relay case, the ASER was analyzed under a proposed beamforming scheme in \cite{6784552}.
    In addition, with the consideration of complex CCI, the distributions of the signal-to-noise ratio (SNR) and ASER were provided in \cite{7308010}.
	In \cite{7230282}, an approximated closed-form expression of EC was derived for an enhanced model $L$ with multi-antenna satellite and multi-antenna user.
	For the multi-user case, the OP performance was analyzed under optimistic user scheduling \cite{7185366} and the NOMA transmission scheme \cite{p512, n45,n39}.
    Specially, to exploit the spectrum efficiency, a novel AF relay mode was applied in \cite{n45,n39}, where the relay not only forwards the signal of the ST but also serves ground users.
	
	Some research efforts have also investigated enhanced model $L$ with multiple relays and multi-hop relays~\cite{7094777}.
	In \cite{7981353}, spectrum sharing between the satellite PU and terrestrial SUs was considered, and the OP of the PU was minimized by selecting the best relay. 
	In \cite{5947821}, a multi-hop AF relay network was analyzed where the maximum ratio combining (MRC) technique was used at receivers. 
	In \cite{7332924}, a relay selection scheme was investigated with multi-antenna satellites.
	In \cite{p262}, the OP performance of a multi-relay multi-user HSTN was analyzed, and the authors presented a relay selection scheme based on the rain attenuation value.
	Recently, in \cite{n70}, the OP performance of a MIMO-enabled HSTN was investigated where the satellite, relay and user are all equipped with multiple antennas. These results have shown potential performance gains from the increasing system cost, i.e., appending antennas and relays. However, current studies still focus on the gain only. Motivated by practical applications, the corresponding cost model should be investigated, and the benefit-to-cost ratio (BCR) should be optimized in future works.

	In addition to the above works, some special scenarios with more practical constraints and new types of transmission regimes have also been discussed. In particular, Arti \textit{et al.} derived the ASER and the average capacity for AF relays based on imperfect CSI \cite{7247422}. Upadhyay \textit{et al.} derived the OP expression with multiple users and an AF relay based on outdated CSI  \cite{7561027}.
	The performance of the two-way relay was investigated in\cite{7918502,n67}. In this regime, there are two sources (the satellite and the user) and one common relay. The two sources transmit to the relay in the first phase, and the relay  simultaneously transmits to both sources in the second phase. 
	Recently, Sharma \textit{et al.} analyzed the OP of an overlay HSTN where multiple IoT receivers work as relays to forward data from the satellite while simultaneously transmitting their own signals using the same frequency \cite{n27}.

	In general, the DF mode outperforms the AF mode due to the decoding gain \cite{6266136,6449258,6655280,7343591}. For the DF mode, An \textit{et al.} investigated the capacity and OP performance of HSTNs \cite{n76, 7098351}. By taking  hardware impairment into account, the OP performance was analyzed in \cite{n69,n38}. Xie \textit{et al.}  investigated the NOMA scheme of HSTNs, and both the AF and DF protocols were considered \cite{n79}. Sharma \textit{et al.} applied multiple UAVs as relays and evaluated the OP performance under opportunistic relay selection \cite{n37}.

	\subsubsection{Resource Management}
	Due to the asymmetric round-trip time (RTT), the resource management in model $L$ is quite different from that of conventional scenarios. We summarize the related literature in Table \ref{table-ResourceManagementL}. In particular, Xu \textit{et al.} proposed a joint relay selection and power allocation scheme \cite{7794885}. Yan \textit{et al.} optimized the beamforming vector of the relay to maximize the secure rate 	\cite{7506598}. By assuming the satellite can communicate with the ST either through a direct link or coordinating with the relay, Ruan \textit{et al.} proposed an adaptive transmission scheme by selecting the transmission mode and optimizing the transmit power  \cite{p513}. 
	Huang \textit{et al.}  investigated the potential gain of the UAV relay and designed a beamforming scheme for energy efficiency maximization \cite{n36}. Ji \textit{et al.} proposed an information forward strategy for remote users and presented a joint resource allocation scheme for different links to improve the energy efficiency of the satellite \cite{n40}. Taking delay requirements into account, Ji \textit{et al.} further proposed an efficient resource allocation scheme for backhaul networks \cite{n6}. 

	\begin{figure}[b]
	\centering
	\includegraphics[width=2.8in]{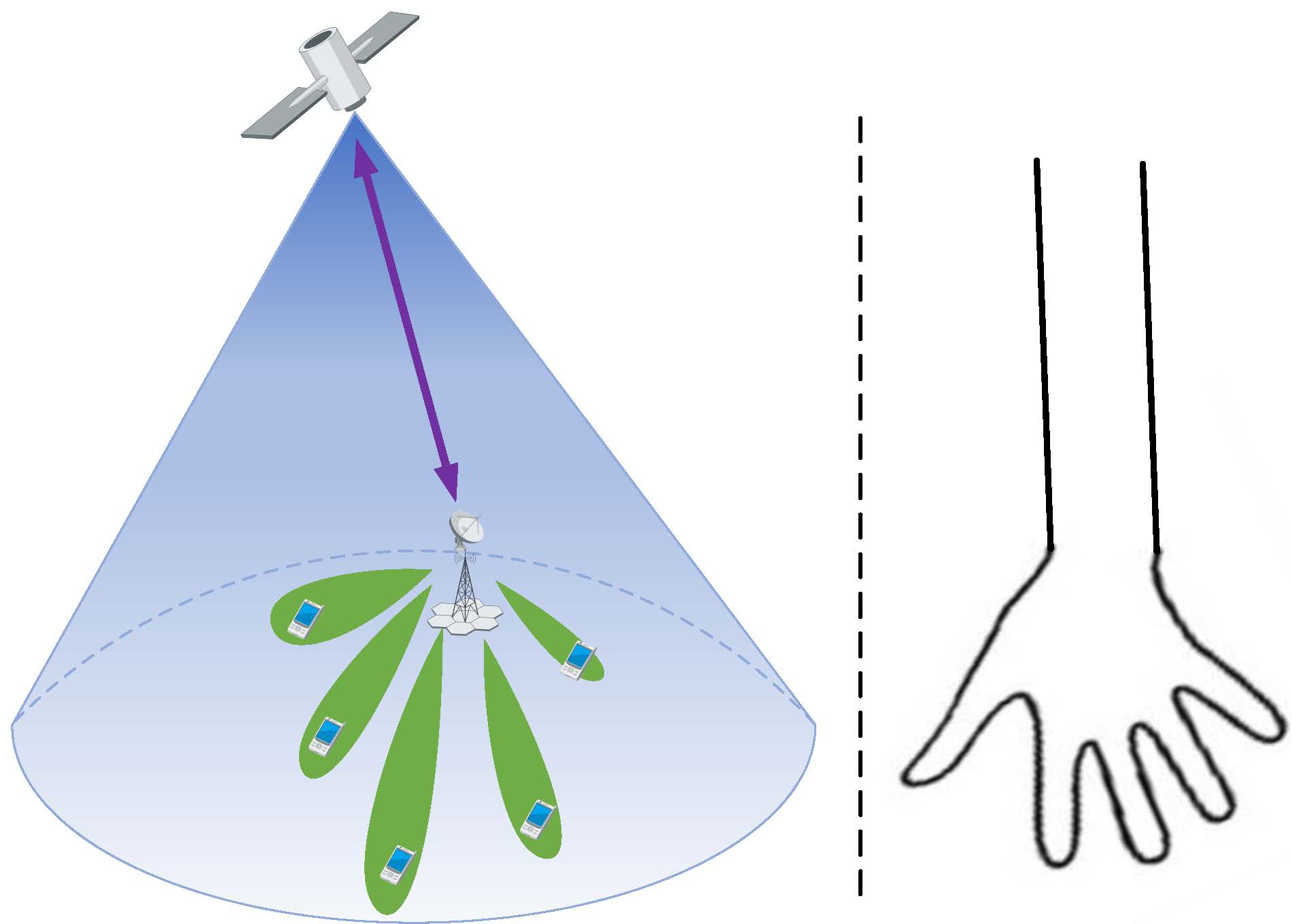}
	\caption{Illustration of an arm-hand-like HSTN, derived from the extension of model $L$.}
	\label{fig_arm}
\end{figure}

	\subsubsection{Security Issue}
	In comparison with terrestrial links, satellite transmissions are much more open due to fewer scatters and longer transmission distances, which inevitably provide opportunities for illegal hackers and lead to security problems. Model $L$  
	could help to tackle this problem through sophisticated relaying designs.  
 	In \cite{7341014}, the achievable secrecy capacity was derived with eavesdroppers and AF relays. The beamforming scheme was discussed in\cite{p259} to maximize the secrecy rate of HSTNs. The security performance with multiple colluding eavesdroppers was analyzed in \cite{p205,n77}, and the case of non-colluding eavesdroppers was discussed in \cite{p273}. For the extension scenarios of model $L$ with multiple relays, in \cite{p236,p239,n16,n46}, the authors addressed the issue of relay selection to improve the secrecy performance. Nevertheless, some important issues including the antenna patterns of satellites/relays, more practical channel models, and more aggressive eavesdropping behaviors, should be considered in future studies to exploit model $L$ and its extensions for security enhancement in practice.  			
			
	\subsubsection{Future Directions}
	Most existing works on model $L$ have considered the satellite-terrestrial difference in channel models. 
	However, only a few studies have paid attention to the difference in the transmission delay. In addition, the processing delay and mobility of MEO/LEO satellites within the relaying duration have not been widely discussed. Note that the dynamic topology of MEO/LEO satellites brings non-negligible handover time, and how to match the delay difference between SatComs and TerComs is still unsolved. Moreover, dynamic beam tracking and adaptive processing at the relay could be further investigated to flexibly adapt to the changing environment. Finally, similar to model $X$, accurate models to characterize the system cost and BCR-oriented optimization frameworks are interesting future research directions for model $L$.
	
	As shown in Fig. \ref{fig_arm}, the extension of model $L$ with multiple terrestrial links forms an arm-hand-like HSTN, where the satellite provides large-scale coverage and beams of the relay achieve high-precision user targeting. 
	Generally, in practice, the relays can be deployed on any mobile platforms, e.g., a car, a vessel, or an airplane, constructing many promising hierarchical network architectures with elastic coverage capabilities. We may mimic the smart synergistic behavior of arms and hands, i.e., the coarse adjustment of arms and the accurate control of hands,  and accordingly create new ways to utilize model $L$. Actually, it is quite interesting to learn from nature, and explore a bionic network research direction, not only for HSTNs, but also for other complicated networks.          

	\subsection{Model $V$}
	
\begin{figure}[t]
	\centering
	\includegraphics[width=2.8in]{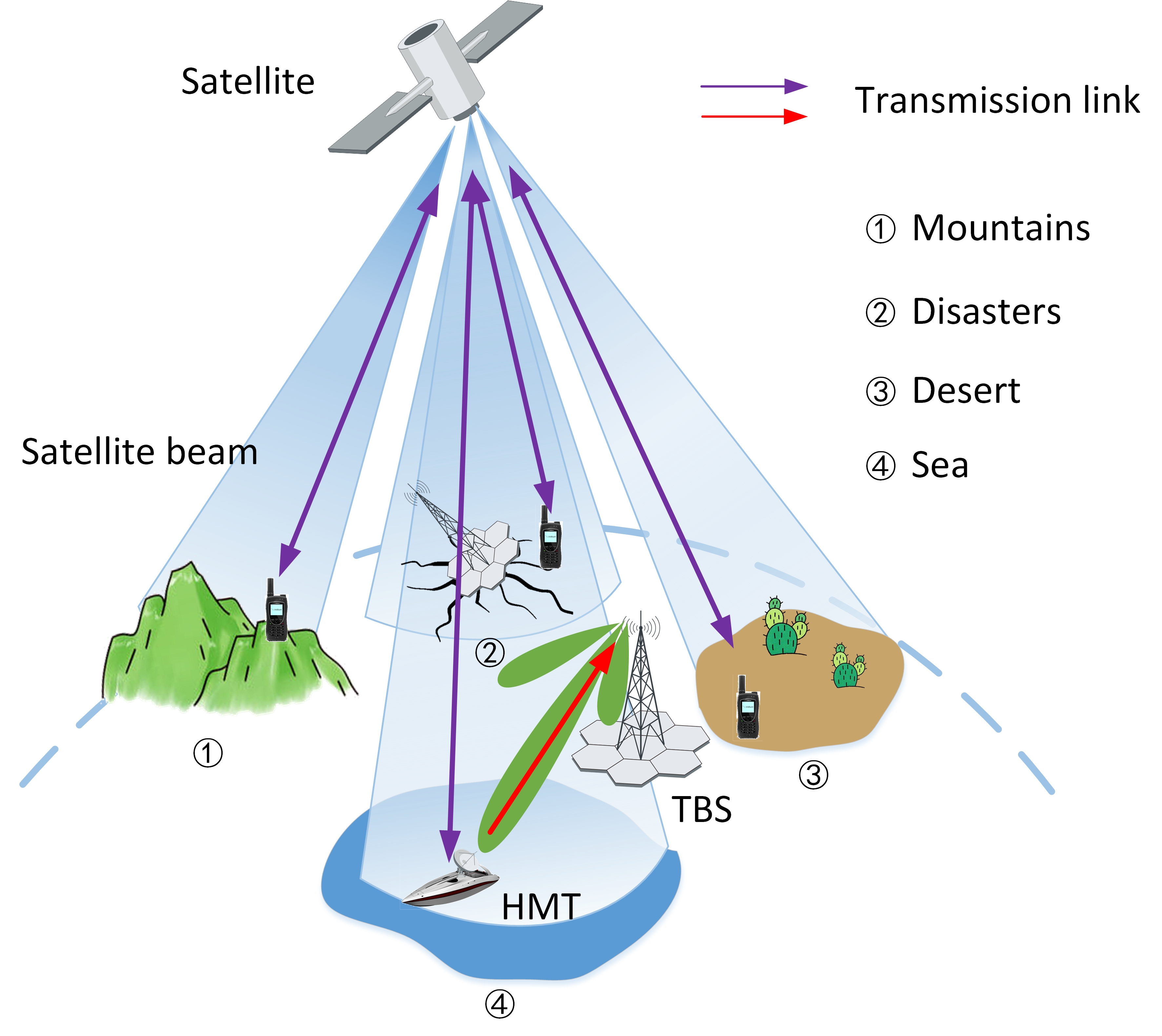}
	\caption{Illustration of the model $V$.}
	\label{fig5}
	\end{figure}

	As shown in Fig. \ref{fig5}, model $V$ consists of one satellite, one BS, and one HMT. When both satellites and TBSs are available, e.g., in urban areas, the HMT mainly connects TBSs for cheaper broadband services. However, in the mountainous, disaster, desert and marine areas, terrestrial facilities usually become unavailable, and the HMT turns to satellites for uninterrupted communication services. In addition to these two either-or fashions, the satellite and the TBS could also work in a coordinated way. One possible example is to use the satellite for low-rate wide-area signaling and use the TBS for high-speed data transmissions with local pencil beams, according to the reported position information by signaling. From the service-management perspective, we may further allocate broadcast and unicast services to satellite and terrestrial links, respectively, to utilize the coverage difference of SatComs and TerComs. In general, the HMT is a two-in-one user, and thus, the acquired QoS is much higher than that of models $X$ and $L$, which provides a new dimension for satellite-terrestrial integration. However, a satellite-terrestrial dual-mode terminal is much more expensive and complex than the dual-/multi-mode terminal of the current cellular networks. There are open problems to be solved regarding efficiently using model $V$ as well. Next, we discuss existing studies on model $V$ from the perspectives of system performance, resource management, and the handover issue.

	\subsubsection{Performance Analysis}
	
	In model $V$, multi-diversity reception is usually used to compensate for the large attenuation of SatComs. At the HMT,
	MRC and selective combining (SC) can be exploited to combine the signals from the satellite and the TBS. In the MRC scheme, the HMT forms a new signal with its carrier-to-noise ratio (CNR) equal to the sum of the CNRs of incoming signals, while in the SC scheme, the HMT selects the best signal for diversity gain.
	The OP performance of MRC and SC was analyzed in \cite{5286328} and \cite{p235}, where the potential gain of MRC compared with SC was shown. Similar to model $X$ and model $L$, the performance of model $V$ under different channel fading conditions and with different system configurations, e.g., the number of antennas equipped on satellites/TBSs  should be further investigated. 
	
	\subsubsection{Resource Management}
	In model $V$, how to choose the appropriate access point, i.e., the satellite and the TBS, and how to allocate radio resources need to be carefully considered.
	Khan \textit{et al.} proposed a multi-radio access algorithm \cite{4536840}.
	Choi \textit{et al.}  investigated a scheduling strategy regarding whether to transmit to the mobile user directly or relay the signal by a ground gateway, and the beamforming, user scheduling, as well as routing were jointly optimized \cite{6785345}.
	Particularly for post-disaster communications, Fujinoin \textit{et al.} introduced a resource allocation method for efficient satellite-terrestrial integration \cite{6050565}. To promote broadcast services, the space-time coding scheme was studied in \cite{5378733,6088980}. The Alamouti space-time code and prefilter were used to mitigate the echoes in HSTNs~\cite{6333071}. Note that more options usually mean more cost. In general, model $V$ requires more complicated system overhead and more expensive hardware. 
	However, until now, there has been few works towards holistically modeling the cost of model $V$ and accordingly investigating BCR-oriented resource allocation strategies.

	\subsubsection{Handover Issue}
	
	\begin{figure}[t]
		\centering
		\includegraphics[width=2.8in]{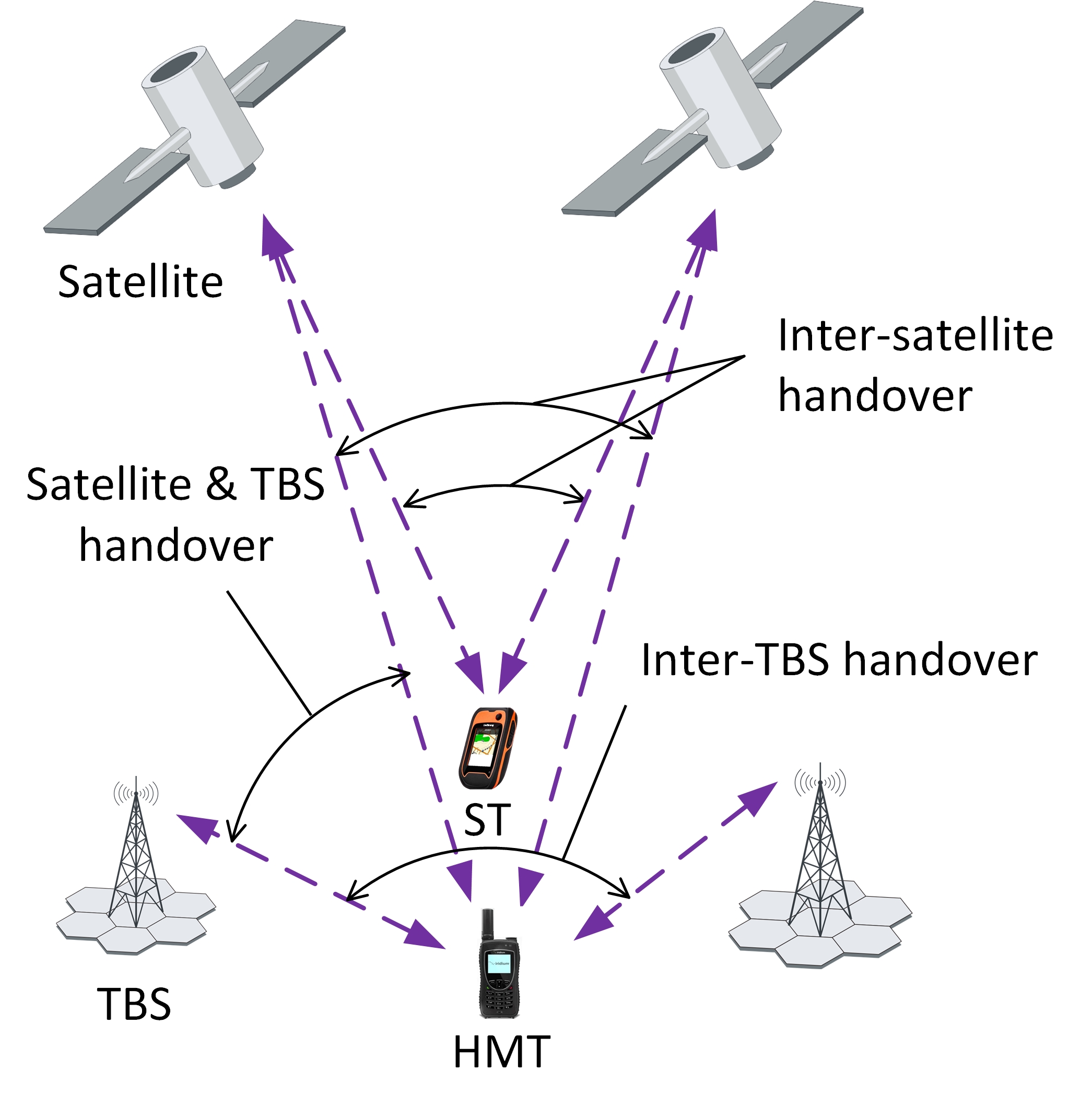}
		\caption{Two kinds of handover for model $V$. }
		\label{fig-handover}
	\end{figure}
	
	As shown in Fig. \ref{fig-handover}, when mobile users or MEO/LEO satellites move, the network topology will change, and the handover is needed between satellites and TBSs. Generally, in model $V$ and its extensions, the handovers can be divided into two types, namely, horizontal and vertical handovers. When a user moves to the edge of the serving satellite or TBS, it is passed on to the other satellite or TBS, which is called the horizontal handover. If the handover occurs between the satellite and TBS, it is regarded as the vertical handover \cite{7463014}. 
	In \cite{1189162}, Yeo \textit{et al.} studied bandwidth allocation and handover management in LEO HSTNs. To achieve global roaming, Akyildiz \textit{et al.} proposed a comprehensive design of the mobility management, where an interworking agent was introduced for both horizontal and vertical handovers \cite{n9}. Crosnier \textit{et al.} studied the handover problem with the satellite as the backhaul \cite{5956545}. Liu \textit{et al.} proposed a named data networking regime for fast handovers \cite{p255}.
	
	Vertical handover often occurs in model $V$ when the user or satellite moves. Sadek \textit{et al.} investigated the handover decision between the GEO satellite and the TBS \cite{6364249}. Kamga \textit{et al.}  studied the handover problem by combining MIMO techniques and proposed an optimal user-driven handoff algorithm \cite{7510941}. Fan \textit{et al.} discussed a suite of signaling protocols, including registration, call setup and inter-segment handover for HMTs to enable Internet Protocol (IP)-based HSTNs \cite{n10}. Considering the different working regimes of ground and space domains, the handover between satellite and terrestrial systems may lead to long delays. In addition, the bidirectional mobility of satellites and users further increases the uncertainty regarding handover issues. In the future, seamless switching needs to be achieved by taking the above two challenges into account. 
		
	\subsubsection{Future Directions}
	It should be noted that the cooperative processing for model $V$ may result in high inter-system communication complexity and additional overhead. In addition, protocol transformation and matching are required because the communication schemes and transmission rates between satellite and terrestrial systems do not match. Based on this, it is necessary to further study low overhead multi-system cooperative interactions, including inter-system information transfer optimization and inter-system rate matching.
	
\begin{figure}[t]
	\centering
	\includegraphics[width=2.8in]{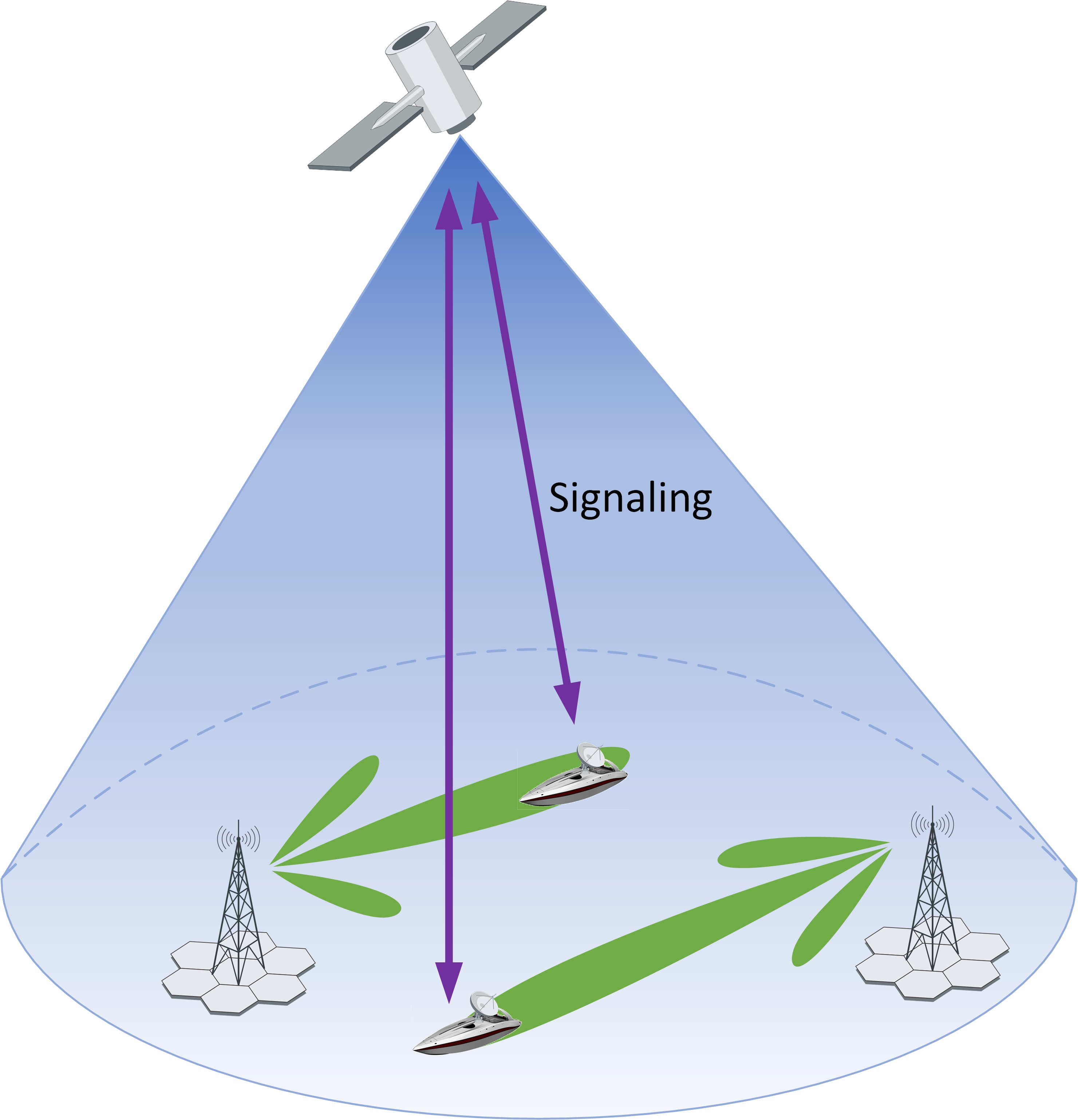}
	\caption{Illustration of a control-communication decoupled HSTN, derived from the extension of model $V$.}
	\label{fig_decouple}
\end{figure}

	As shown in Fig. \ref{fig_decouple}, the extension of model $V$ may produce a control-communication coordinated HSTN, where the satellite provides wide-area signaling, and the TBS adopts pencil beams to efficiently serve target users according to the information, e.g., the positions of users, provided by the narrowband control subsystem. This framework may greatly improve the efficiency of the terrestrial subsystem since it no longer needs to cover the whole area, and thus is able to focus radio resources on the target users. When the requirement occurs in the blind areas of TBSs, the satellite control subsystem will report this demand at once, and a pencil beam will be dispatched to the corresponding user. Nevertheless, due to the differences between SatComs and TerComs, especially in terms of latency, how to achieve timely interactions between the two subsystems is still challenging. 	
	A promising approach for overcoming this problem is to use the extrinsic information, e.g., the shipping lane information of marine users, and historical information, e.g., the communication behavior of users, to establish an integrated on-demand and prediction-driven response regime, which however still remains open.      

	\section{Open Issues}
	To date, the 3GPP has finished all standardization works of NTNs in 3GPP Release 16. New normative solutions of 5G new radio (NR) in NTNs are under investigation in 3GPP Release 17, and a long-term study in Release 18 as well as Release 19 is being carried out \cite{n22}. Although new 5G infrastructures are currently being deployed, which would bring excellent performance improvements, many challenges still exist to meet the increasing requirements of future ubiquitous IoT. The seamless, low delay and high capacity demands all call for the establishment of an agile, smart, and secure HSTN. To this end, more research efforts should focus on deriving the fundamental theories and technologies for each basic cooperative model. On that basis, the intelligent orchestration of models $X$, $L$, and $V$ is important to create a HSTN in an on-demand manner. In addition, some frontier technologies, such as artificial intelligence, MEC, and blockchain technologies, can be leveraged. These technologies are under rapid development and become new promoters of the smart integration of SatComs and TerComs. Next, we briefly outline these potential open issues.
	
	{\subsection{BCR Maximization for Basic Cooperative Models}
	Since the establishment of a HSTN is usually costly, we have to carefully evaluate its BCR, rather than only considering its gains. To this end, a holistic HSTN  cost model  should be studied, and the multidimensional gains should be mathematically characterized. On that basis, the BCR optimization framework could be established for each basic cooperative model.    
	
	\subsection{Intelligent Orchestration of Basic Cooperative Models}
	The proposed three basic cooperative models provide a mesoscopic sight to fill the gap between micro link analysis and macro network evaluation of HSTNs. To satisfy varying user demands, a practical HSTN should be capable of dynamically changing the combination fashion of basic cooperative models, and we call this intelligent orchestration of basic cooperative models. To do so, a cyber agent can be introduced to gather data on service requirements, environmental information, and network status, intelligently deciding whether and how to change the orchestration. Among these procedures, the theoretical bounds and foundational trade-offs should be given for  practical guidance. In addition, the cyber agent is expected to abstract knowledge from historical behaviors, upgrade with the network and become increasingly intelligent. Towards this end,  a knowledge-driven network architecture may be established for HSTNs.  

	\subsection{Interplay Between HSTNs and Other Technologies}
	Machine learning technologies, e.g., deep reinforcement learning, can be utilized to promote HSTNs, especially to solve the hard-to-model problems therein. On the other hand, the differences between SatComs and TerComs also require the upgrading of conventional machine learning methods, e.g., federated learning in HSTNs should be redesigned to adapt to the network conditions.
	
	{
	Smart caching and MEC can be adopted into HSTNs, leading to a caching, computing and communication integrated network. The differences among caching, computing and communication, coupled with the differences between SatComs and TerComs, will bring about huge complexity. Systematic methods and economic methodologies, e.g., smart pricing, can be used to tackle this problem.}
	
	Blockchain technology can be utilized to establish an open ecology for resource allocation in HSTNs. In addition to reducing the complexity of network management, it may also achieve increased efficiency and security. Combined with SatComs, the inherent  broadcasting merit
	of satellites can also help to increase the
	efficiency, e.g., transactions per second, of traditional blockchain systems.} 
	
	\section{Conclusions}
	
	In this paper, we have proposed three basic cooperative models, i.e., model $X$, model $L$, and model $V$, to better understand large-scale heterogeneous HSTNs. The differences between SatComs and TerComs have been summarized, and the state-of-the-art technologies for each model have been reviewed. We have shown possible future directions towards establishing a cell-free, hierarchical, decoupled HSTN. We have also outlined open issues to envision an agile, smart, and secure HSTN for the upcoming 6G ubiquitous IoT.


\begin{thebibliography}{100}
	\providecommand{\url}[1]{#1}
	\csname url@samestyle\endcsname
	\providecommand{\newblock}{\relax}
	\providecommand{\bibinfo}[2]{#2}
	\providecommand{\BIBentrySTDinterwordspacing}{\spaceskip=0pt\relax}
	\providecommand{\BIBentryALTinterwordstretchfactor}{4}
	\providecommand{\BIBentryALTinterwordspacing}{\spaceskip=\fontdimen2\font plus
		\BIBentryALTinterwordstretchfactor\fontdimen3\font minus
		\fontdimen4\font\relax}
	\providecommand{\BIBforeignlanguage}[2]{{%
			\expandafter\ifx\csname l@#1\endcsname\relax
			\typeout{** WARNING: IEEEtran.bst: No hyphenation pattern has been}%
			\typeout{** loaded for the language `#1'. Using the pattern for}%
			\typeout{** the default language instead.}%
			\else
			\language=\csname l@#1\endcsname
			\fi
			#2}}
	\providecommand{\BIBdecl}{\relax}
	\BIBdecl
	\bibitem{n211}
	M. {Pan}, P. {Li} and Y. {Fang}, ``Cooperative communication aware link scheduling for cognitive vehicular networks,'' \emph{IEEE J. Sel. Areas Commun.}, vol. 30, no. 4, pp. 760-768, May 2012.
	
	\bibitem{n89}
	N.~{Cao}, Y.~{Chen}, X.~{Gu} \emph{et~al.}, ``Joint bi-static radar and
	communications designs for intelligent transportation,'' \emph{IEEE Trans.
		Veh. Tech.}, vol.~69, no.~11, pp. 13\,060--13\,071, Nov. 2020.
	
	\bibitem{n90}
	N.~{Cao}, Y.~{Chen}, X.~{Gu} \emph{et~al.}, ``Joint radar-communication
	waveform designs using signals from multiplexed users,'' \emph{IEEE Trans.
		Commun.}, vol.~68, no.~8, pp. 5216--5227, Aug. 2020.
	
	\bibitem{n91}
	R.~{Girau}, M.~{Anedda}, M.~{Fadda} \emph{et~al.}, ``{Coastal} monitoring
	system based on social {Internet} of {Things} platform,'' \emph{IEEE Internet
		Things J.}, vol.~7, no.~2, pp. 1260--1272, 2020.
	
	\bibitem{Wei2021}
	T.~{Wei}, W.~{Feng}, Y.~{Chen} \emph{et~al.}, ``{Hybrid satellite-terrestrial
		communication networks for the maritime Internet of Things: Key technologies,
		opportunities, and challenges},'' \emph{IEEE Internet Things J.}, 2021, Early
	Access.
	
	\bibitem{n92}
	A.~{Salam} and S.~{Shah}, ``{Internet} of {Things} in smart agriculture:
	{Enabling} technologies,'' in \emph{2019 IEEE 5th World Forum on Internet of
		Things (WF-IoT)}, 2019, pp. 692--695.
	
	\bibitem{n93}
	H.~{Lu}, Q.~{Liu}, D.~{Tian} \emph{et~al.}, ``The cognitive internet of
	vehicles for autonomous driving,'' \emph{IEEE Netw.}, vol.~33, no.~3, pp.
	65--73, 2019.
	
	\bibitem{n94}
	M.~A. {Mahmud}, K.~{Bates}, T.~{Wood} \emph{et~al.}, ``A complete {Internet} of
	{Things} ({IoT}) platform for structural health monitoring ({SHM}),'' in
	\emph{2018 IEEE 4th World Forum Internet Things (WF-IoT)}, 2018, pp.
	275--279.
	
	\bibitem{n95}
	J.~J. {Wellington} and P.~{Ramesh}, ``Role of {Internet} of {Things} in
	disaster management,'' in \emph{2017 Int. Conf. Innov. Inf., Embedded Commun.
		Syst. (ICIIECS)}, 2017, pp. 1--4.
	
	\bibitem{n96}
	J.~{Chen}, W.~{Feng}, J.~{Xing} \emph{et~al.}, ``Hybrid beamforming/combining
	for millimeter wave {MIMO}: A machine learning approach,'' \emph{IEEE Trans.
		Veh. Tech.}, vol.~69, no.~10, pp. 11\,353--11\,368, 2020.
	
	\bibitem{p4002}
	W.~{Feng}, J.~{Wang}, Y.~{Chen} \emph{et~al.}, ``{UAV}-aided {MIMO}
	communications for {5G} {Internet} of {Things},'' \emph{IEEE Internet Things
		J.}, vol.~6, no.~2, pp. 1731--1740, Apr. 2019.
	
	\bibitem{n97}
	C.~{Liu}, W.~{Feng}, Y.~{Chen} \emph{et~al.}, ``Cell-free satellite-{UAV}
	networks for {6G} wide-area {Internet} of {Things},'' \emph{IEEE J. Sel.
		Areas Commun.}, pp. 1--1, 2020.
	
	\bibitem{n98}
	H.~{Wei}, W.~{Feng}, C.~{Zhang} \emph{et~al.}, ``Creating efficient blockchains
	for the {Internet} of {Things} by coordinated satellite-terrestrial
	networks,'' \emph{IEEE Wireless Commun.}, vol.~27, no.~3, pp. 104--110, Jun.
	2020.
	
	\bibitem{p4001}
	X.~{Li}, W.~{Feng}, J.~{Wang} \emph{et~al.}, ``Enabling {5G} on the ocean: {A}
	hybrid satellite-{UAV}-terrestrial network solution,'' \emph{IEEE Wireless
		Commun.}, vol.~27, no.~6, pp. 116--121, Dec. 2020.
	
	\bibitem{p253}  
	A. Kapovits, M. {Corici}, I. {Gheorghe-Pop} \textit{et al.}, ``Satellite communications integration with terrestrial networks," \textit{China Commun.}, vol. 15, no. 8, pp. 22--38, Aug. 2018.
	

	
	\bibitem{5247781}
	J.~K. Chamberlain and R.~G. Medhurst, ``Mutual interference between communication satellites and terrestrial line-of-sight radio-relay systems," in \emph{Proc. Institution Electrical Engineers}, vol. 111, no.~3, pp. 524--534, Mar. 1964.

	\bibitem{5247517}
	J.~K. Chamberlain, ``Interference between an earth station of a communication-satellite system and the stations of terrestrial line-of-sight radio-relay systems," in \emph{Proc. Institution Electrical Engineers}, vol. 112, no.~2, pp. 231--241, Feb. 1965.
	
	\bibitem{4206709}
	P.~{Johns}, ``Interference between terrestrial line-of-sight radio-relay
	systems and communication-satellite systems,'' \emph{Electronics Lett.},
	vol.~2, no.~5, pp. 177--178, May 1966.
	
	\bibitem{1145889}
	J.~{Lee}, ``Symbiosis between a terrestrial-based integrated services digital
	network and a digital satellite network,'' \emph{IEEE J. Sel. Areas Commun.},
	vol.~1, no.~1, pp. 103--109, Jan. 1983.
	
	\bibitem{10414}
	M.~{Richharia} and B.~{Evans}, ``Synergy between land mobile satellite and
	terrestrial systems-possibilities in the {European} region,'' in \emph{Proc.
		Institution Electrical Engineers}, no.~2, Oct. 1988, pp. 111--117.
	
	\bibitem{166759}
	C.~Caini, G.~E. Corazza, G.~Falciasecca \emph{et al.},  ``A spectrum-and-power-efficient EHF mobile satellite system to be integrated with terrestrial cellular systems," \emph{IEEE J. Sel. Areas Commun.}, vol.~10, no.~8, pp. 1315--1325, Oct. 1992.
	
	\bibitem{672027}
	T.~{Hatsuda}, ``Theoretical and measuring results of interferences between
	fixed-satellite and terrestrial radio network in urban and rural areas at 4
	{GHz} and 11 {GHz} bands,'' in \emph{Proc. AMPC Asia-Pacific Microwave Conf},
	no.~2, Adelaide, Australia, Aug. 1992, pp. 255--258.
	
	\bibitem{222474}
	S.~{Liu}, ``Satellite and terrestrial microwave communications in {China},''
	\emph{IEEE Commun. Mag.}, vol.~31, no.~7, pp. 38--40, Jul. 1993.
	
	\bibitem{497076}
	F.~{Ananasso} and F.~{Priscoli}, ``Satellite personal communications:
	{Integration} scenarios with terrestrial networks,'' in \emph{Proc. 4th IEEE
		Intern. Conf. Universal Personal Commun. (ICUPC)}, Tokyo, Japan, Nov. 1995,
	pp. 585--589.
	
	\bibitem{864080}
	C.~W. {Bond} and C.~{Curran}, ``The integration of personal satellite
	communication with terrestrial cellular: {A} business perspective,'' in
	\emph{Proc. 1996 European Workshop Mobile/Personal Satcoms (EMPS)}, Rome,
	Italy, 1996, pp. 501--507.
	
	\bibitem{p275}
	\BIBentryALTinterwordspacing
	{3rd Generation Partnership Project}, ``Release 16,'' {Accessed}: Sep. 16,
	2020. [Online]. Available: \url{https://www.3gpp.org/release-16}.
	\BIBentrySTDinterwordspacing
	
	\bibitem{p248}
	B.~T. {Jou}, O.~{Vidal}, J.~{Cahill} \emph{et~al.}, ``Architecture options for
	satellite integration into {5G} networks,'' in \emph{Proc. 2018 Euro. Conf.
		Netw. \& Commun. (EuCNC)}, Ljubljana, Slovenia, 2018, pp. 398--399.
	
	\bibitem{n2}
	\BIBentryALTinterwordspacing
	{SaT5G Project}, ``Sa{T5G} progress and main achievements,'' {Accessed}: Sep.
	16, 2020. [Online]. Available:
	\url{https://www.sat5g-project.eu/wp-content/uploads/2020/05/0-SaT5G_ReviewMeeting2-PM-update.pdf}.
	\BIBentrySTDinterwordspacing
	
	\bibitem{p276}
	\BIBentryALTinterwordspacing
	{SaT5G Project}, ``Sat{NE}x {IV} satellite network of experts {IV},''
	{Accessed}: Sep. 16, 2020. [Online]. Available:
	\url{https://artes.esa.int/projects/satnex-iv}.
	\BIBentrySTDinterwordspacing
	
	\bibitem{7245586}
	R.~M. {Burkhart}, ``Tutorial: Propagation phenomena and terrestrial
	interference in satellite television transmission,'' \emph{SMPTE J.},
	vol.~98, no.~9, pp. 658--663, Sep. 1989.
	
	\bibitem{5714025}
	T.~{Taleb}, Y.~{Hadjadj-Aoul}, and T.~{Ahmed}, ``Challenges, opportunities, and
	solutions for converged satellite and terrestrial networks,'' \emph{IEEE
		Wireless Commun.}, vol.~18, no.~1, pp. 46--52, Feb. 2011.
	
	\bibitem{7463014}
	C.~{Niephaus}, M.~{Kretschmer}, and G.~{Ghinea}, ``Qo{S} provisioning in
	converged satellite and terrestrial networks: {A} survey of the
	state-of-the-art,'' \emph{IEEE Commun. Surveys Tuts.}, vol.~18, no.~4, pp.
	2415--2441, 4th Quart. 2016.
	
	\bibitem{p269}
	J.~{Liu}, Y.~{Shi}, Z.~M. {Fadlullah} \emph{et~al.}, ``Space-air-ground
	integrated network: {A} survey,'' \emph{IEEE Commun. Surveys Tuts.}, vol.~20,
	no.~4, pp. 2714--2741, May 2018.
	
	\bibitem{n3}
	P.~{Wang}, J.~{Zhang}, X.~{Zhang} \emph{et~al.}, ``Convergence of satellite and
	terrestrial networks: {A} comprehensive survey,'' \emph{IEEE Access}, vol.~8,
	pp. 5550--5588, Dec. 2020.
	
	\bibitem{p274}
	H.~{Yao}, L.~{Wang}, X.~{Wang} \emph{et~al.}, ``The space-terrestrial
	integrated network: {An} overview,'' \emph{IEEE Commun. Mag.}, vol.~56,
	no.~9, pp. 178--185, Sep. 2018.
	
	\bibitem{p240}
	R.~{Gopal} and N.~{BenAmmar}, ``Framework for unifying {5G} and next generation
	satellite communications,'' \emph{IEEE Netw.}, vol.~32, no.~5, pp. 16--24,
	Sep./Oct. 2018.
	
	\bibitem{p241}
	G.~{Giambene}, S.~{Kota}, and P.~{Pillai}, ``Satellite-{5G} integration: {A}
	network perspective,'' \emph{IEEE Netw.}, vol.~32, no.~5, pp. 25--31, Sep.
	2018.
	
	\bibitem{p252}
	C.~{Dai}, G.~{Zheng}, and Q.~{Chen}, ``Satellite constellation design with
	multi-objective genetic algorithm for regional terrestrial satellite
	network,'' \emph{China Commun.}, vol.~15, no.~8, pp. 1--10, Aug. 2018.
	
	\bibitem{p220}
	B.~{Di}, H.~{Zhang}, L.~{Song} \emph{et~al.}, ``Ultra-dense {LEO}:
	{Integrating} terrestrial-satellite networks into {5G} and beyond for data
	offloading,'' \emph{IEEE Trans. Wireless Commun.}, vol.~18, no.~1, pp.
	47--62, Jan. 2019.
	
	\bibitem{1391449}
	P.~{Pace}, G.~{Aloi}, F.~{De Rango} \emph{et~al.}, ``An integrated
	satellite-{HAP}-terrestrial system architecture: {Resources} allocation and
	traffic management issues,'' in \emph{Proc. 2004 IEEE Veh. Tech. Conf.},
	vol.~5, Milan, Italy, May 2004, pp. 2872--2875.
	
	\bibitem{n99}
	X.~{Li}, W.~{Feng}, Y.~{Chen} \emph{et~al.}, ``Maritime coverage enhancement
	using {UAVs} coordinated with hybrid satellite-terrestrial networks,''
	\emph{IEEE Trans. Commun.}, vol.~68, no.~4, pp. 2355--2369, Apr. 2020.
	
	\bibitem{p217}
	Y.~{Shi}, Y.~{Cao}, J.~{Liu} \emph{et~al.}, ``{A cross-domain SDN architecture
		for multi-layered space-terrestrial integrated networks},'' \emph{IEEE
		Netw.}, vol.~33, no.~1, pp. 29--35, Jan. 2019.
	
	\bibitem{p263}
	Y.~{Shi}, J.~{Liu}, Z.~M. {Fadlullah} \emph{et~al.}, ``Cross-layer data
	delivery in satellite-aerial-terrestrial communication,'' \emph{IEEE Wireless
		Commun.}, vol.~25, no.~3, pp. 138--143, Jun. 2018.
	
	\bibitem{5264626}
	W.~R. {Stone}, \emph{Mobile, Terrestrial, and Satellite Propagation
		Modeling}.\hskip 1em plus 0.5em minus 0.4em\relax Wiley-IEEE Press, 1999.
	
	\bibitem{6493416}
	S.~A. {Kanellopoulos}, A.~D. {Panagopoulos}, C.~I. {Kourogiorgas}
	\emph{et~al.}, ``Satellite and terrestrial links rain attenuation time series
	generator for heavy rain climatic regions,'' \emph{IEEE Trans. Antennas
		Propag.}, vol.~61, no.~6, pp. 3396--3399, Jun. 2013.
	
	\bibitem{1137538}
	K.~I. {Timothy}, {J. Ong}, and E.~B.~L. {Choo}, ``Raindrop size
	distribution using method of moments for terrestrial and satellite
	communication applications in {Singapore},'' \emph{IEEE Trans. Antennas
		Propag.}, vol.~50, no.~10, pp. 1420--1424, Oct. 2002.
	
	\bibitem{6772645}
	L.~T. {Gusler} and D.~C. {Hogg}, ``{Some calculations on coupling between
		satellite communications and terrestrial radio-relay systems due to
		scattering by rain},'' \emph{Bell Sys. Tech. J.}, vol.~49, no.~7, pp.
	1491--1511, Sep. 1970.
	
	\bibitem{6883690}
	G.~N. {Kamga}, M.~{Xia}, and S.~{Aïssa}, ``Unified {MIMO} channel model for
	mobile satellite systems with ancillary terrestrial component,'' in
	\emph{Proc. 2014 IEEE Int. Conf. Commun. (ICC)}, Sydney, Australia, Jun.
	2014, pp. 2449--2453.
	
	\bibitem{p3008}
	B.~{Wang}, C.~{Huang}, J.~{Xu} \emph{et~al.}, ``Predicting wireless {mmWave}
	massive {MIMO} channel characteristics using machine learning algorithms,''
	\emph{Wireless Commun. Mob. Com.}, vol. 2018, Aug. 2018.
	
	\bibitem{p3007}
	L.~{Bai}, C.~{Wang}, Q.~{Xu} \emph{et~al.}, ``{Prediction of channel excess
		attenuation for satellite communication systems at Q-band using artificial
		neural network},'' \emph{IEEE Antennas Wireless Propaga. Lett.}, vol.~18,
	no.~11, pp. 2235--2239, Nov. 2019.
	
	\bibitem{p501}
	S.~K. {Sharma}, S.~{Chatzinotas}, and B.~{Ottersten}, ``{Satellite cognitive
		communications: Interference modeling and techniques selection},'' in
	\emph{Proc. 2012 6th Advanced Satell. Multimedia Sys. Conf. (ASMS) and 12th
		Signal Processing Space Commun. Workshop (SPSC)}, Sep. 2012, pp. 111--118.
	
	\bibitem{p502}
	S.~K. {Sharma}, S.~{Chatzinotas}, and B.~{Ottersten}, ``Cognitive radio
	techniques for satellite communication systems,'' in \emph{Proc. 2013 IEEE
		78th Veh. Tech. Conf. (VTC)}, Las Vegas, USA, Sep. 2013, pp. 1--5.
	
	\bibitem{7417497}
	J.~{Janhunen}, J.~{Ketonen}, A.~{Hulkkonen} \emph{et~al.}, ``Satellite uplink
	transmission with terrestrial network interference,'' in \emph{Proc. 2015
		IEEE Global Commun. Conf. (GLOBECOM)}, San Diego, CA, USA, 2015, pp. 1--6.
	
	\bibitem{p504}
	S.~{Maleki}, S.~{Chatzinotas}, J.~{Krause} \emph{et~al.}, ``{Cognitive zone for
		broadband satellite communications in 17.3-17.7 GHz band},'' \emph{IEEE
		Wireless Commun. Lett.}, vol.~4, no.~3, pp. 305--308, Jun. 2015.
	
	\bibitem{7343438}
	X.~{Zhu}, R.~{Shi}, W.~{Feng} \emph{et~al.}, ``Position-assisted interference
	coordination for integrated terrestrial-satellite networks,'' in \emph{Proc.
		2015 IEEE Int. Symp. Personal, Indoor \& Mobile Radio Commun. (PIMRC)}, Hong
	Kong, China, 2015, pp. 971--975.
	
	\bibitem{6166516}
	A.~H. {Khan}, M.~A. {Imran}, and B.~G. {Evans}, ``Iterative turbo beamforming
	for orthogonal frequency division multiplexing-based hybrid
	terrestrial-satellite mobile system,'' \emph{IET commun.}, vol.~6, no.~2, pp.
	157--164, Jan. 2012.
	
	
	\bibitem{640258}
	T.~{Hatsuda} and Y.~{Motozumi}, ``Interference experiments between
	fixed-satellite and terrestrial radio-relay services,'' \emph{IEEE Trans.
		Aerosp. Electron. Syst.}, vol.~34, no.~1, pp. 23--32, Jan. 1998.
	
	\bibitem{7556323}
	H.~{Li}, H.~{Yin}, F.~{Dong} \emph{et~al.}, ``Capacity upper bound analysis of
	the hybrid satellite terrestrial communication systems,'' \emph{IEEE Commun.
		Lett.}, vol.~20, no.~12, pp. 2402--2405, Dec. 2016.
	
	\bibitem{7878033}
	H.~{Li}, Q.~{Yao}, Y.~{He} \emph{et~al.}, ``Performance analysis of the return
	link for the hybrid satellite terrestrial communication systems,'' in
	\emph{Proc. 2016 IEEE Int. Conf. Signal Process. (ICSP)}, Chengdu, China,
	Nov. 2016, pp. 1283--1287.
	
	\bibitem{7510840}
	K.~{An}, M.~{Lin}, T.~{Liang} \emph{et~al.}, ``On the ergodic capacity of
	multiple antenna cognitive satellite terrestrial networks,'' in \emph{Proc.
		2016 IEEE Int. Conf. Commun. (ICC)}, Kuala Lumpur, Malaysia, May 2016, pp.
	1--5.
	
	\bibitem{7996553}
	K.~{An}, M.~{Lin}, J.~{Guyang} \emph{et~al.}, ``Outage performance for the
	cognitive broadband satellite system and terrestrial cellular network in
	millimeter wave scenario,'' in \emph{Proc. 2017 IEEE Int. Conf. Commun.
		(ICC)}, Paris, France, May 2017, pp. 1--6.
	
	\bibitem{n51}
	X.~{Yan}, K.~{An}, T.~{Liang} \emph{et~al.}, ``Effect of imperfect channel
	estimation on the performance of cognitive satellite terrestrial networks,''
	\emph{IEEE Access}, vol.~7, pp. 126\,293--126\,304, Sep. 2019.
	
	\bibitem{p514}
	X. Yan, H. Xiao, C. Wang \emph{et al.}, ``On the ergodic capacity of NOMA-based cognitive hybrid satellite terrestrial networks," in \textit{Proc. IEEE/CIC Int. Conf. Commun. China}, Qingdao, China, Oct. 2017, pp.~1--5.
	
	
	\bibitem{p515}
	Y. Ruan, Y. Li, C. Wang \emph{et al.}, ``Effective capacity analysis for underlay cognitive satellite-terrestrial networks,"		in \textit{Proc. IEEE Intern. Conf. Commun.}, Paris, France, May 2017, pp.~1--6.
	
	\bibitem{7248442}
	G.~N. {Kamga}, M.~{Xia}, and S.~{Aissa}, ``A unified performance evaluation of
	integrated mobile satellite systems with ancillary terrestrial component,''
	in \emph{Proc. 2015 IEEE Int. Conf. Commun. (ICC)}, London, UK, 2015, pp.934--938.
	
	\bibitem{7898463}
	Y.~{Ruan}, Y.~{Li}, C.~{Wang} \emph{et~al.}, ``Outage performance of integrated
	satellite-terrestrial networks with hybrid CCI,'' \emph{IEEE Commun. Lett.},
	vol.~21, no.~7, pp. 1545--1548, Jul. 2017.
	

	

	
	\bibitem{n75}
	K.~{An}, T.~{Liang}, G.~{Zheng} \emph{et~al.}, ``{Performance limits of
		cognitive uplink FSS and terrestrial FS for Ka-band},'' \emph{IEEE Trans.
		Aerosp. Electron. Syst.}, vol.~55, no.~5, pp. 2604--2611, Oct. 2019.
	
	\bibitem{6517359}
	S.~{Vassaki}, M.~I. {Poulakis}, A.~D. {Panagopoulos} \emph{et~al.}, ``{Power
		allocation in cognitive satellite terrestrial networks with QoS
		constraints},'' \emph{IEEE Commun. Lett.}, vol.~17, no.~7, pp. 1344--1347,
	Jul. 2013.
	
	\bibitem{7391092}
	E.~{Lagunas}, S.~K. {Sharma}, S.~{Maleki} \emph{et~al.}, ``{Power control for
		satellite uplink and terrestrial fixed-service co-existence in Ka-band},'' in
	\emph{Proc. 2015 IEEE Veh. Tech. Conf. (VTC)}, Sep. 2015, pp. 1--5.
	
	\bibitem{7373246}
	U.~{Park}, H.~W. {Kim}, D.~S. {Oh} \emph{et~al.}, ``{Performance analysis of
		dynamic resource allocation for interference mitigation in integrated
		satellite and terrestrial systems},'' in \emph{Proc. 2015 9th Int. Conf. Next
		Gener. Mobile Appl., Services Tech.}, Cambridge, UK, 2015, pp. 217--221.
	
	\bibitem{p237}
	B.~{Gao}, M.~{Lin}, K.~{An} \emph{et~al.}, ``{ADMM-based optimal power control
		for cognitive satellite terrestrial uplink networks},'' \emph{IEEE Access},
	vol.~6, pp. 64\,757--64\,765, Oct. 2018.
	
	\bibitem{7882651}
	S.~{Shi}, G.~{Li}, K.~{An} \emph{et~al.}, ``Optimal power control for real-time
	applications in cognitive satellite terrestrial networks,'' \emph{IEEE
		Commun. Lett.}, vol.~21, no.~8, pp. 1815--1818, Aug. 2017.
	
	\bibitem{n41}
	Z.~{Chen}, D.~{Guo}, G.~{Ding} \emph{et~al.}, ``{Optimized power control scheme
		for global throughput of cognitive satellite-terrestrial networks based on
		non-cooperative game},'' \emph{IEEE Access}, vol.~7, pp. 81\,652--81\,663,
	Jun. 2019.
	
	\bibitem{n71}
	J.~{Hu}, G.~{Li}, D.~{Bian} \emph{et~al.}, ``{Optimal power control for
		cognitive LEO constellation with terrestrial networks},'' \emph{IEEE Commun.
		Lett.}, vol.~24, no.~3, pp. 622--625, Mar. 2020.
	
	\bibitem{n78}
	J.~{Wang}, B.~{Zhang}, L.~{Jia} \emph{et~al.}, ``{A distributed collaborative
		game-theoretic approach in cognitive satellite communication networks},''
	\emph{IEEE Access}, vol.~8, pp. 129\,446--129\,460, Jul. 2020.
	
	\bibitem{p208}
	M.~{Hua}, Y.~{Wang}, M.~{Lin} \emph{et~al.}, ``{Joint CoMP transmission for
		UAV-aided cognitive satellite terrestrial networks},'' \emph{IEEE Access},
	vol.~7, pp. 14\,959--14\,968, Jan. 2019.
	
	\bibitem{p511}
	Y.~{Ruan}, Y.~{Li}, C.~{Wang} \emph{et~al.}, ``{Power allocation in cognitive
		satellite-vehicular networks from energy-spectral efficiency tradeoff
		perspective},'' \emph{IEEE Trans. Cogn. Commun. Netw.}, vol.~5, no.~2, pp.
	318--329, Jun. 2019.
	
	\bibitem{p3001}
	Y.~{Ruan}, Y.~{Li}, C.~{Wang} \emph{et~al.}, ``Energy efficient power
	allocation for delay constrained cognitive satellite terrestrial networks
	under interference constraints,'' \emph{IEEE Trans. Wireless Commun.},
	vol.~18, no.~10, pp. 4957--4969, Oct. 2019.
	
	\bibitem{n4}
	Y.~{Ruan}, L.~{Jiang}, Y.~{Li} \emph{et~al.}, ``{Energy-efficient power control
		for cognitive satellite-terrestrial networks with outdated CSI},'' \emph{IEEE
		Syst. J.}, pp. 1--4, 2020, Early Access.
	
	\bibitem{5586925}
	V.~{Deslandes}, J.~{Tronc}, and A.~{Beylot}, ``Analysis of interference issues
	in integrated satellite and terrestrial mobile systems,'' in \emph{Proc. 2010 5th Advanced Satell. Multimedia Sys. Conf. \& the 11th Signal Process. Space Commun. Workshop}, 2010, pp. 256--261.
	
	\bibitem{7248431}
	W.~{Tang}, P.~{Thompson}, and B.~{Evans}, ``{Frequency sharing between
		satellite and terrestrial systems in the Ka band: A database approach},'' in
	\emph{Proc. 2015 IEEE Int. Conf. Commun.(ICC)}, London, UK, Jun. 2015, pp.
	867--872.
	
	\bibitem{7928993}
	W.~{Feng}, N.~{Ge}, and J.~{Lu}, ``{Coordinated satellite-terrestrial networks:
		A robust spectrum sharing perspective},'' in \emph{Proc. 2017 26th Wireless
		Opt. Commun. Conf. (WOCC)}, Newark, NJ, USA, Apr. 2017, pp. 1--5.
	
	\bibitem{7986309}
	X.~{Zhu}, C.~{Jiang}, W.~{Feng} \emph{et~al.}, ``{Resource allocation in
		spectrum-sharing cloud based integrated terrestrial-satellite network},'' in
	\emph{Proc. 2017 13th Int. Wireless Commun. Mobile Comput. Conf. (IWCMC)},
	Jun. 2017, pp. 334--339.
	
	\bibitem{p260}
	M.~{Jia}, X.~{Zhang}, X.~{Gu} \emph{et~al.}, ``{Energy efficient cognitive
		spectrum sharing scheme based on inter-cell fairness for integrated
		satellite-terrestrial communication systems},'' in \emph{Proc. 2018 87th IEEE
		Veh. Tech. Conf. (VTC)}, Porto, Portugal, Jun. 2018, pp. 1--6.
	
	\bibitem{p219}
	W.~{Wang}, S.~{Zhao}, Y.~{Zheng} \emph{et~al.}, ``Resource allocation method of
	cognitive satellite terrestrial networks under non-ideal spectrum sensing,''
	\emph{IEEE Access}, vol.~7, pp. 7957--7964, Jan. 2019.
	
	\bibitem{7336495}
	E.~{Lagunas}, S.~K. {Sharma}, S.~{Maleki} \emph{et~al.}, ``{Resource allocation
		for cognitive satellite communications with incumbent terrestrial
		networks},'' \emph{IEEE Trans. Cognitive Commun. Netw.}, vol.~1, no.~3, pp.
	305--317, Sep. 2015.
	
	\bibitem{7962743}
	E.~{Lagunas}, S.~{Maleki}, {L. Lei} \emph{et~al.}, ``{Carrier allocation for
		hybrid satellite-terrestrial backhaul networks},'' in \emph{Proc. 2017 IEEE
		Int. Conf. Commun. Workshops (ICC)}, Paris, France, May 2017, pp. 718--723.
	
	\bibitem{n42}
	Z.~{Chen}, D.~{Guo}, K.~{An} \emph{et~al.}, ``{Efficient and fair resource
		allocation scheme for cognitive satellite-terrestrial networks},'' \emph{IEEE
		Access}, vol.~7, pp. 145\,124--145\,133, Sep. 2019.
	
	\bibitem{n83}
	I.~{Ahmed}, H.~{Khammari}, A.~{Shahid} \emph{et~al.}, ``{A survey on hybrid
		beamforming techniques in 5G: Architecture and system model perspectives},''
	\emph{IEEE Commun. Surveys Tuts.}, vol.~20, no.~4, pp. 3060--3097, 4th Quart.
	2018.
	
	\bibitem{5286394}
	A.~H. {Khan}, M.~A. {Imran}, and B.~G. {Evans}, ``{OFDM based adaptive
		beamforming for hybrid terrestrial-satellite mobile system with pilot
		reallocation},'' in \emph{Proc. 2009 Int. Workshop Satell. Space Commun.},
	Tuscany, Italy, Sep. 2009, pp. 201--205.
	
	\bibitem{6294500}
	A.~H. {Khan}, M.~A. {Imran}, and B.~G. {Evans}, ``{Semi-adaptive beamforming
		for OFDM based hybrid terrestrial-satellite mobile system},'' \emph{IEEE
		Trans. Wireless Commun.}, vol.~11, no.~10, pp. 3424--3433, Oct. 2012.
	
	\bibitem{7058570}
	K.~A. {Shah}, {W. Yong}, W.~{Ur Rehman} \emph{et~al.}, ``{Simulation of
		elimination of co-channel interference in hybrid terrestrial-satellite mobile
		communication system using adaptive beam-forming technique},'' in \emph{Proc.
		2015 12th Int. Bhurban Conf. Appl. Sci. Tech. (IBCAST)}, Islamabad, Pakistan,
	2015, pp. 620--622.
	
	\bibitem{6636830}
	S.~K. {Sharma}, S.~{Chatzinotas}, and B.~{Ottersten}, ``{Transmit beamforming
		for spectral coexistence of satellite and terrestrial networks},'' in
	\emph{Proc. 8th Int. Conf. Cogn. Radio Oriented Wireless Netw.}, DC, USA,
	Jul. 2013, pp. 275--281.
	
	\bibitem{p506}
	S.~K. {Sharma}, S.~{Maleki}, S.~{Chatzinotas} \emph{et~al.}, ``{Joint carrier
		allocation and beamforming for cognitive SatComs in Ka-band (17.3-18.1
		GHz)},'' in \emph{Proc. 2015 IEEE Int. Conf. Commun. (ICC)}, London, UK, Jun.
	2015, pp. 873--878.
	
	\bibitem{7980751}
	X.~{Artiga}, M. Á. {Vázquez} ,  A. {Pérez-Neira}  \emph{et~al.}, ``{Spectrum
		sharing in hybrid terrestrial-satellite backhaul networks in the Ka band},''
	in \emph{Proc. 2017 European Conf. Netw. Commun. (EuCNC)}, Oulu, Finland,
	Jun. 2017, pp. 1--5.
	
	\bibitem{p267}
	S.~{Vuppala}, M.~{Sellathurai}, and S.~{Chatzinotas}, ``{Optimal deployment of
		base stations in cognitive satellite-terrestrial networks},'' in \emph{Proc.
		2018 22th Int. ITG Workshop Smart Antennas}, Bochum, Germany, Mar. 2018, pp.
	1--8.
	
	\bibitem{p232}
	Y. {Jiang}, J.~{Ouyang}, C. {Yin} \emph{et~al.}, ``{Downlink
		beamforming scheme for hybrid satellite-terrestrial networks},'' \emph{IET
		Commun.}, vol.~12, no.~18, pp. 2342--2346, Nov. 2018.
	
	\bibitem{p268}
	Z.~{Lin}, M.~{Lin}, J.~{Ouyang} \emph{et~al.}, ``{Beamforming for secure
		wireless information and power transfer in terrestrial networks coexisting
		with satellite networks},'' \emph{IEEE Signal Process. Lett.}, vol.~25,
	no.~8, pp. 1166--1170, Aug. 2018.
	
	\bibitem{p272}
	M.~{Lin}, Z.~{Lin}, W.~{Zhu} \emph{et~al.}, ``{Joint beamforming for secure
		communication in cognitive satellite terrestrial networks},'' \emph{IEEE J.
		Sel. Areas Commun.}, vol.~36, no.~5, pp. 1017--1029, May 2018.
	
	\bibitem{n43}
	Z.~{Lin}, M.~{Lin}, J.~{Wang} \emph{et~al.}, ``{Joint beamforming and power
		allocation for satellite-terrestrial integrated networks with nonorthogonal
		multiple access},'' \emph{IEEE J. Sel. Areas Commun.}, vol.~13, no.~3, pp.
	657--670, Jun. 2019.
	
	\bibitem{n64}
	Q.~{Zhang}, K.~{An}, X.~{Yan} \emph{et~al.}, ``{Coexistence and performance
		limits for the cognitive broadband satellite system and mmWave cellular
		network},'' \emph{IEEE Access}, vol.~8, pp. 51\,905--51\,917, Mar. 2020.
	
	\bibitem{n58}
	Y.~{Ruan}, Y.~{Li}, R.~{Zhang} \emph{et~al.}, ``{Cooperative resource
		management for cognitive satellite-aerial-terrestrial integrated networks
		towards IoT},'' \emph{IEEE Access}, vol.~8, pp. 35\,759--35\,769, Feb. 2020.
	
	\bibitem{n5}
	C.~{Liu}, W.~{Feng}, Y.~{Chen} \emph{et~al.}, ``{Optimal beamforming for hybrid
		satellite terrestrial networks with nonlinear PA and imperfect CSIT},''
	\emph{IEEE Wireless Commun. Lett.}, vol.~9, no.~3, pp. 276--280, Mar. 2020.
	
	\bibitem{n100}
	T.~{Wei}, W.~{Feng}, N.~{Ge} \emph{et~al.}, ``{Environment-aware coverage
		optimization for space-ground integrated maritime communications},''
	\emph{IEEE Access}, vol.~8, pp. 89\,205--89\,214, May 2020.
	
	\bibitem{p204}
	A.~{Guidotti}, A.~{Vanelli-Coralli}, M.~{Conti} \emph{et~al.}, ``{Architectures
		and key technical challenges for 5G systems incorporating satellites},''
	\emph{IEEE Trans. Veh. Tech.}, vol.~68, no.~3, pp. 2624--2639, 2019.
	
	\bibitem{p211}
	W.~{Chien}, C.~{Lai}, M.~S. {Hossain} \emph{et~al.}, ``{Heterogeneous space and
		terrestrial integrated networks for IoT: Architecture and challenges},''
	\emph{IEEE Netw.}, vol.~33, no.~1, pp. 15--21, Jan. 2019.
	
	\bibitem{n88}
	T.~{Huang}, W.~{Yang}, J.~{Wu} \emph{et~al.}, ``{A survey on green 6G network:
		Architecture and technologies},'' \emph{IEEE Access}, vol.~7, pp.
	175\,758--175\,768, Dec. 2019.
	
	\bibitem{n86}
	G.~{Charbit}, D.~{Lin}, K.~{Medles} \emph{et~al.}, ``{Space-terrestrial radio
		network integration for IoT},'' in \emph{Proc. 2020 2nd 6G Wireless Summit
		(6G SUMMIT)}, Levi, Finland, 2020, pp. 1--5.
	
	\bibitem{n85}
	J. Liu, X. Du, J. Cui \emph{et al.}, ``Task-oriented intelligent networking architecture for the space-air-ground-aqua integrated network," \textit{IEEE Internet Things J.}, vol. 7, no. 6, pp. 5345--5358, Jun. 2020.
	
	\bibitem{n23}
	L.~{Bertaux}, S.~{Medjiah}, P.~{Berthou} \emph{et~al.}, ``{Software defined
		networking and virtualization for broadband satellite networks},'' \emph{IEEE
		Commun. Mag.}, vol.~53, no.~3, pp. 54--60, Mar. 2015.
	
	\bibitem{p254}
	L.~{Boero}, M.~{Marchese}, and F.~{Patrone}, ``{The impact of delay in
		software-defined integrated terrestrial-satellite networks},'' \emph{China
		Commun.}, vol.~15, no.~8, pp. 11--21, Aug. 2018.
	
	\bibitem{p215}
	K.~{Lin}, D.~{Wang}, L.~{Hu} \emph{et~al.}, ``{Virtualized QoS-driven spectrum
		allocation in space-terrestrial integrated networks},'' \emph{IEEE Netw.},
	vol.~33, no.~1, pp. 58--63, Jan. 2019.
	
	\bibitem{p249}
	C.~{Niephaus}, J.~{Mödeker}, and G.~{Ghinea}, ``{Toward traffic offload in
		converged satellite and terrestrial networks},'' \emph{IEEE Trans.
		Broadcasting}, vol.~65, no.~2, pp. 340--346, Jun. 2019.
	
	\bibitem{p209}
	Z.~{Zhang}, W.~{Zhang}, and F.~{Tseng}, ``{Satellite mobile edge computing:
		Improving QoS of high-speed satellite-terrestrial networks using edge
		computing techniques},'' \emph{IEEE Netw.}, vol.~33, no.~1, pp. 70--76, Jan.
	2019.
	
	\bibitem{p216}
	Y.~{Bi}, G.~{Han}, S.~{Xu} \emph{et~al.}, ``{Software defined space-terrestrial
		integrated networks: Architecture, challenges, and solutions},'' \emph{IEEE
		Netw.}, vol.~33, no.~1, pp. 22--28, Jan. 2019.
	
	\bibitem{n102}
	B.~{Feng}, H.~{Zhou}, H.~{Zhang} \emph{et~al.}, ``{HetNet: A flexible
		architecture for heterogeneous satellite-terrestrial networks},'' \emph{IEEE
		Netw.}, vol.~31, no.~6, pp. 86--92, Nov./Dec. 2017.
	
	\bibitem{n103}
	B.~{Feng}, G.~{Li}, G.~{Li} \emph{et~al.}, ``{Enabling efficient service
		function chains at terrestrial-satellite hybrid cloud networks},'' \emph{IEEE
		Netw.}, vol.~33, no.~6, pp. 94--99, Nov./Dec. 2019.
	
	\bibitem{n82}
	A.~{Papa}, T.~{de Cola}, P.~{Vizarreta} \emph{et~al.}, ``{Design and evaluation
		of reconfigurable SDN LEO constellations},'' \emph{IEEE Trans. Netw. Service
		Manag.}, vol.~17, no.~3, pp. 1432--1445, Sep. 2020.
	
	\bibitem{n80}
	D.~{Chen}, C.~{Yang}, P.~{Gong} \emph{et~al.}, ``{Resource cube: Multi-virtual
		resource management for integrated satellite-terrestrial industrial IoT
		networks},'' \emph{IEEE Trans. Veh. Technol.}, vol.~69, no.~10, pp.
	11\,963--11\,974, Oct. 2020.
	
	\bibitem{n84}
	G.~{Wang}, S.~{Zhou}, S.~{Zhang} \emph{et~al.}, ``{SFC-based service
		provisioning for reconfigurable space-air-ground integrated networks},''
	\emph{IEEE J. Sel. Areas Commun.}, vol.~38, no.~7, pp. 1478--1489, Jul. 2020.
	
	\bibitem{n104}
	B.~{Feng}, Z.~{Cui}, Y.~{Huang} \emph{et~al.}, ``{Elastic resilience for
		software-defined satellite networking: Challenges, solutions, and open
		issues},'' \emph{IT Professional}, vol.~22, no.~6, pp. 39--45, Nov./Dec.
	2020.
	
	\bibitem{5586888}
	A.~{Iqbal} and K.~M. {Ahmed}, ``{SER analysis of cooperative
		satellite-terrestrial network over non identical fading channels},'' in
	\emph{Proc. 2010 5th Advanced Satell. Multimedia Syst. Conf. \& 11th Signal
		Processing Space Commun. Workshop}, Cagliari, Italy, Sep. 2010, pp. 329--334.
	
	\bibitem{7504407}
	Y.~{Ruan}, Y.~{Li}, R.~{Zhang} \emph{et~al.}, ``{Performance analysis of hybrid
		satellite-terrestrial cooperative networks with distributed Alamouti code},''
	in \emph{Proc. 2016 IEEE Veh. Tech. Conf. (VTC)}, Nanjing, China, May 2016,
	pp. 1--5.
	
	\bibitem{6589295}
	M.~R. {Bhatnagar} and M.~K. {Arti}, ``{Performance analysis of AF based hybrid
		satellite-terrestrial cooperative network over generalized fading
		channels},'' \emph{IEEE Commun. Lett.}, vol.~17, no.~10, pp. 1912--1915, Oct.
	2013.
	
	\bibitem{6605500}
	M.~R. {Bhatnagar} and M.~K. {Arti}, ``{Performance analysis of hybrid
		satellite-terrestrial FSO cooperative system},'' \emph{IEEE Photonics Tech.
		Lett.}, vol.~25, no.~22, pp. 2197--2200, Nov. 2013.
	
	\bibitem{6784552}
	M.~K. {Arti} and M.~R. {Bhatnagar}, ``{Beamforming and combining in hybrid
		satellite-terrestrial cooperative systems},'' \emph{IEEE Commun. Lett.},
	vol.~18, no.~3, pp. 483--486, Mar. 2014.
	
	\bibitem{7308010}
	L.~{Yang} and M.~O. {Hasna}, ``{Performance analysis of amplify-and-forward
		hybrid satellite-terrestrial networks with cochannel interference},''
	\emph{IEEE Trans. Commun.}, vol.~63, no.~12, pp. 5052--5061, Dec. 2015.
	
	\bibitem{7230282}
	K.~{An}, M.~{Lin}, T.~{Liang} \emph{et~al.}, ``{Performance analysis of
		multi-antenna hybrid satellite-terrestrial relay networks in the presence of
		interference},'' \emph{IEEE Trans. Commun.}, vol.~63, no.~11, pp. 4390--4404,
	Nov. 2015.
	
	\bibitem{7185366}
	K.~{An}, M.~{Lin}, and T.~{Liang}, ``{On the performance of multiuser hybrid
		satellite-terrestrial relay networks with opportunistic scheduling},''
	\emph{IEEE Commun. Lett.}, vol.~19, no.~10, pp. 1722--1725, Oct. 2015.
	
	\bibitem{p512}
	X.~{Yan}, H.~{Xiao}, C.~{Wang} \emph{et~al.}, ``{Outage performance of
		NOMA-based hybrid satellite-terrestrial relay networks},'' \emph{IEEE
		Wireless Commun. Lett.}, vol.~7, no.~4, pp. 538--541, Aug. 2018.
	
	\bibitem{n45}
	X.~{Zhang}, B.~{Zhang}, K.~{An} \emph{et~al.}, ``{Outage performance of
		NOMA-based cognitive hybrid satellite-terrestrial overlay networks by
		amplify-and-forward protocols},'' \emph{IEEE Access}, vol.~7, pp.
	85\,372--85\,381, Jun. 2019.
	
	\bibitem{n39}
	V.~{Singh}, P.~K. {Upadhyay}, and M.~{Lin}, ``{On the performance of
		NOMA-assisted overlay multiuser cognitive satellite-terrestrial networks},''
	\emph{IEEE Wireless Commun. Lett.}, vol.~9, no.~5, pp. 638--642, May 2020.
	
	\bibitem{7094777}
	M.~Lin, J.~Ouyang, and W. Zhu, ``{On the performance of hybrid
		satellite-terrestrial cooperative networks with interferences},'' in
	\emph{Proc. 2014 48th Asilomar Conf. Signals, Syst. Comput.}, Pacific Grove,
	CA, USA, Nov. 2014, pp. 1796--1800.
	
	\bibitem{7981353}
	P.~K. {Sharma}, P.~K. {Upadhyay}, D.~B. {da Costa} \emph{et~al.},
	``{Performance analysis of overlay spectrum sharing in hybrid
		satellite-terrestrial systems with secondary network selection},'' \emph{IEEE
		Trans. Wireless Commun.}, vol.~16, no.~10, pp. 6586--6601, Oct. 2017.
	
	\bibitem{5947821}
	A.~{Iqbal} and K.~M. {Ahmed}, ``{Outage probability analysis of multi-hop
		cooperative satellite-terrestrial network},'' in \emph{Proc. 2011 8th Electr.
		Eng./Electron., Comput., Telecommun. \& Info. Tech. (ECTI) Association
		Thailand Conf.}, Khon Kaen, Thailand, May 2011, pp. 256--259.
	
	\bibitem{7332924}
	P.~K. {Upadhyay} and P.~K. {Sharma}, ``{Max-max user-relay selection scheme in
		multiuser and multirelay hybrid satellite-terrestrial relay systems},''
	\emph{IEEE Commun. Lett.}, vol.~20, no.~2, pp. 268--271, Feb. 2016.
	
	\bibitem{p262}
	X.~{Liang}, J.~{Jiao}, S.~{Wu} \emph{et~al.}, ``{Outage analysis of multirelay
		multiuser hybrid satellite-terrestrial millimeter-wave networks},''
	\emph{IEEE Wireless Commun. Lett.}, vol.~7, no.~6, pp. 1046--1049, Dec. 2018.
	
	\bibitem{n70}
	A.~{H. G. Swalem}, {J. V. M. Halim}, and {H. Elhennawy}, ``Performance analysis
	of MIMO AF CDMA hybrid satellite-terrestrial cooperative networks using
	multiple relays strategy for downlink,'' \emph{IET Commun.}, vol.~13, pp.
	2155--2162, Aug. 2019.
	
	\bibitem{7247422}
	A.~{M. K.}, ``{Imperfect CSI based AF relaying in hybrid satellite-terrestrial
		cooperative communication systems},'' in \emph{Proc. 2015 IEEE Int. Conf.
		Commun. Workshop (ICCW)}, London, UK, Jun. 2015, pp. 1681--1686.
	
	\bibitem{7561027}
	P.~K. {Upadhyay} and P.~K. {Sharma}, ``{Multiuser hybrid satellite-terrestrial
		relay networks with co-channel interference and feedback latency},'' in
	\emph{Proc. 2016 European Conf. Netw. Commun. (EuCNC)}, Athens, Greece, Jun.
	2016, pp. 174--178.
	
	\bibitem{7918502}
	K.~{Guo}, B.~{Zhang}, Y.~{Huang} \emph{et~al.}, ``{Performance analysis of
		two-way satellite terrestrial relay networks with hardware impairments},''
	\emph{IEEE Wireless Commun. Lett.}, vol.~6, no.~4, pp. 430--433, Aug. 2017.
	
	\bibitem{n67}
	W.~{Zeng}, J.~{Zhang}, D.~W.~K. {Ng} \emph{et~al.}, ``{Two-way hybrid
		terrestrial-satellite relaying systems: Performance analysis and relay
		selection},'' \emph{IEEE Trans. Veh. Technol.}, vol.~68, no.~7, pp.
	7011--7023, Jul. 2019.
	
	\bibitem{n27}
	{P. K. Sharma}, B.~{Yogesh}, D.~{Gupta} \emph{et~al.}, ``{Overlay
		satellite-terrestrial networks for IoT under hybrid interference
		environments},'' Mar. 2020, \url{Arxiv: 2003.12950}.
	
	\bibitem{6266136}
	S.~{Sreng}, B.~{Escrig}, and M.~{Boucheret}, ``{Outage analysis of hybrid
		satellite-terrestrial cooperative network with best relay selection},'' in
	\emph{Proc. 2012 Wireless Telecommun. Symp.}, London, UK, Apr. 2012, pp.
	1--5.
	
	\bibitem{6449258}
	S.~{Sreng}, B.~{Escrig}, and M.~{Boucheret}, ``{Exact symbol error probability
		of hybrid/integrated satellite-terrestrial cooperative network},'' \emph{IEEE
		Trans. Wireless Commun.}, vol.~12, no.~3, pp. 1310--1319, Mar. 2013.
	
	\bibitem{6655280}
	S.~{Sreng}, B.~{Escrig}, and M.~{Boucheret}, ``{Exact outage probability of a
		hybrid satellite terrestrial cooperative system with best relay selection},''
	in \emph{Proc. 2013 IEEE Int. Conf. Commun. (ICC)}, Budapest, Hungary, Jun.
	2013, pp. 4520--4524.
	
	\bibitem{7343591}
	Y.~{Zhao}, L.~{Xie}, H.~{Chen} \emph{et~al.}, ``{Ergodic channel capacity
		analysis of the hybrid satellite-terrestrial single frequency network},'' in
	\emph{Proc. 2015 IEEE 26th Annual Int. Symp. Pers., Indoor, Mobile Radio
		Commun. (PIMRC)}, Hong Kong, China, Aug. 2015, pp. 1803--1807.
	
	\bibitem{n76}
	K.~{An} and T.~{Liang}, ``{Hybrid satellite-terrestrial relay networks with
		adaptive transmission},'' \emph{IEEE Trans. Veh. Technol.}, vol.~68, no.~12,
	pp. 12\,448--12\,452, Dec. 2019.
	
	\bibitem{7098351}
	K.~{An}, J.~{Ouyang}, M.~{Lin} \emph{et~al.}, ``{Outage analysis of
		multi-antenna cognitive hybrid satellite-terrestrial relay networks with
		beamforming},'' \emph{IEEE Commun. Lett.}, vol.~19, no.~7, pp. 1157--1160,
	Jul. 2015.
	
	\bibitem{n69}
	X.~{Tang}, K.~{An}, K.~{Guo} \emph{et~al.}, ``{Outage analysis of
		non-orthogonal multiple access-based integrated satellite-terrestrial relay
		networks with hardware impairments},'' \emph{IEEE Access}, vol.~7, pp.
	141\,258--141\,267, Sep. 2019.
	
	\bibitem{n38}
	K.~{Guo}, M.~{Lin}, B.~{Zhang} \emph{et~al.}, ``{Performance analysis of hybrid
		satellite-terrestrial cooperative networks with relay selection},''
	\emph{IEEE Trans. Veh. Technol.}, vol.~69, no.~8, pp. 9053--9067, Aug. 2020.
	
	\bibitem{n79}
	S.~{Xie}, B.~{Zhang}, D.~{Guo} \emph{et~al.}, ``{Performance analysis and power
		allocation for NOMA-based hybrid satellite-terrestrial relay networks with
		imperfect channel state information},'' \emph{IEEE Access}, vol.~7, pp.
	136\,279--136\,289, Sep. 2019.
	
	\bibitem{n37}
	P.~K. {Sharma}, D.~{Deepthi}, and D.~I. {Kim}, ``{Outage probability of 3-D
		mobile UAV relaying for hybrid satellite-terrestrial networks},'' \emph{IEEE
		Commun. Lett}, vol.~24, no.~2, pp. 418--422, Feb. 2020.
	
	\bibitem{7794885}
	Y.~{Xu}, Y.~{Wang}, R.~{Sun} \emph{et~al.}, ``{Joint relay selection and power
		allocation for maximum energy efficiency in hybrid
		satellite-aerial-terrestrial systems},'' in \emph{Proc. 2016 IEEE 27th Int.
		Symp. Pers., Indoor Mobile Radio Commun. (PIMRC)}, Valencia, Spain, Sep.
	2016, pp. 1--6.
	
	\bibitem{7506598}
	Y.~{Yan}, {B. Zhang}, {D. Guo} \emph{et~al.}, ``{Joint beamforming and jamming design for secure cooperative hybrid satellite-terrestrial relay
		network},'' in \emph{Proc. 2016 25th Wireless Optical Commun. Conf. (WOCC)},
	Chengdu, China, May 2016, pp. 1--5.
	
		\bibitem{p513}
	Y. Ruan, Y. Li, C. Wang \emph{et~al.},
	``Energy efficient adaptive transmissions in integrated satellite-terrestrial networks with SER constraints,"		\textit{IEEE Trans. Wireless Commun.}, vol. 17, no. 1, pp.~210--222, Jan. 2018.
	
	\bibitem{n36}
	Q.~{Huang}, M.~{Lin}, J.~{Wang} \emph{et~al.}, ``{Energy efficient beamforming
		schemes for satellite-aerial-terrestrial networks},'' \emph{IEEE Trans.
		Commun.}, vol.~68, no.~6, pp. 3863--3875, Jun. 2020.
	
	\bibitem{n40}
	Z.~{Ji}, S.~{Wu}, C.~{Jiang} \emph{et~al.}, ``{Energy-efficient data offloading
		for multi-cell satellite-terrestrial networks},'' \emph{IEEE Commun. Lett.},
	vol.~24, no.~10, pp. 2265--2269, Oct. 2020.
	
	\bibitem{n6}
	Z.~{Ji}, S.~{Cao}, S.~{Wu} \emph{et~al.}, ``{Delay-aware satellite-terrestrial
		backhauling for heterogeneous small cell networks},'' \emph{IEEE Access},
	vol.~8, pp. 112\,190--112\,202, Jun. 2020.
	
	\bibitem{7341014}
	K.~{An}, M.~{Lin}, T.~{Liang} \emph{et~al.}, ``{Secure transmission in
		multi-antenna hybrid satellite-terrestrial relay networks in the presence of
		eavesdropper},'' in \emph{Proc. 2015 Int. Conf. Wireless Commun. Signal
		Process. (WCSP)}, Nanjing, China, Oct. 2015, pp. 1--5.
	
	\bibitem{p259}
	C.~{Chen} and L.~{Song}, ``{Secure communications in hybrid cooperative
		satellite-terrestrial networks},'' in \emph{Proc. 2018 IEEE 87th Veh. Tech.
		Conf. (VTC)}, Porto, Portugal, Jun. 2018, pp. 1--5.
	
	\bibitem{p205}
	V.~{Bankey} and P.~K. {Upadhyay}, ``{Physical layer security of multiuser
		multirelay hybrid satellite-terrestrial relay networks},'' \emph{IEEE Trans.
		Veh. Tech.}, vol.~68, no.~3, pp. 2488--2501, Mar. 2019.
	
	\bibitem{n77}
	V.~{Bankey} and P.~K. {Upadhyay}, ``{Physical layer security of hybrid
		satellite terrestrial relay networks with multiple colluding eavesdroppers
		over non-identically distributed Nakagami-m fading channels},'' \emph{IET
		Commun.}, vol.~13, pp. 2115--2123, Aug. 2019.
	
	\bibitem{p273}
	V.~{Bankey}, P.~K. {Upadhyay}, D.~B. {da Costa} \emph{et~al.}, ``{Performance
		analysis of multi-antenna multiuser hybrid satellite-terrestrial relay
		systems for mobile services delivery},'' \emph{IEEE Access}, vol.~6, pp.
	24\,729--24\,745, Apr. 2018.
	
	\bibitem{p236}
	W.~{Cao}, Y.~{Zou}, Z.~{Yang} \emph{et~al.}, ``{Relay selection for improving
		physical-layer security in hybrid satellite-terrestrial relay networks},''
	\emph{IEEE Access}, vol.~6, pp. 65\,275--65\,285, Oct. 2018.
	
	\bibitem{p239}
	K.~{Guo}, K.~{An}, B.~{Zhang} \emph{et~al.}, ``{Physical layer security for
		hybrid satellite terrestrial relay networks with joint relay selection and
		user scheduling},'' \emph{IEEE Access}, vol.~6, pp. 55\,815--55\,827, Oct.
	2018.
	
	\bibitem{n16}
	P.~K. {Sharma} and D.~I. {Kim}, ``{Secure 3D mobile UAV relaying for hybrid
		satellite-terrestrial networks},'' \emph{IEEE Trans. Wireless Commun.},
	vol.~19, no.~4, pp. 2770--2784, Apr. 2020.
	
	\bibitem{n46}
	R.~{Xu}, X.~{Da}, H.~{Hu} \emph{et~al.}, ``{A secure hybrid
		satellite-terrestrial communication network with AF/DF and relay
		selection},'' \emph{IEEE Access}, vol.~7, pp. 171\,980--171\,994, Nov. 2019.
	
	\bibitem{5286328}
	D.~{Skraparlis}, V.~K. {Sakarellos}, A.~D. {Panagopoulos} \emph{et~al.},
	``{Satellite and terrestrial diversity reception performance in tropical
		regions},'' in \emph{Proc. 2009 Int. Workshop Satell. Space Commun.},
	Tuscany, Italy, Sep. 2009, pp. 403--406.
	
	\bibitem{p235}
	V.~{Singh}, S.~{Solanki}, and P.~K. {Upadhyay}, ``{Cognitive relaying
		cooperation in satellite-terrestrial systems with multiuser diversity},''
	\emph{IEEE Access}, vol.~6, pp. 65\,539--65\,547, Oct. 2018.
	
	\bibitem{4536840}
	A.~A. {Khan}, M.~{Adda}, and T.~A. {Khan}, ``{Multi radio diversity for
		satellite-terrestrial mobile communications},'' in \emph{Proc. 2008 4th IEEE
		Int. Conf. Circuits Syst. Commun.}, Shanghai, China, May 2008, pp. 673--677.
	
	\bibitem{6785345}
	J.~P. {Choi} and {C. Joo}, ``{Cross-layer optimization for
		satellite-terrestrial heterogeneous networks},'' in \emph{Proc. 2014 Int.
		Conf. Comput., Netw. Commun. (ICNC)}, Honolulu, HI, USA, Feb. 2014, pp.
	276--281.
	
		\bibitem{6050565}
	Y.~{Fujino}, A.~{Miura}, N.~{Hamamoto} \emph{et~al.}, ``Satellite terrestrial
	integrated mobile communication system as a disaster countermeasure,'' in
	\emph{Proc. 2011 URSI General Assembly \& Scientific Symp}, Istanbul, Turkey,
	Aug. 2011, pp. 1--4.
	
	\bibitem{5378733}
	Y.~{Nasser} and J.~{Helard}, ``{Double layer space-time block code for hybrid
		satellite-terrestrial broadcasting systems},'' in \emph{Proc. 2009 IEEE 70th
		Veh. Tech. Conf. (VTC)}, Anchorage, AK, USA, Sep. 2009, pp. 1--5.
	
	\bibitem{6088980}
	C.~{Hollanti}, R.~{Vehkalahti}, and Y.~{Nasser}, ``{Algebraic hybrid
		satellite-terrestrial space-time codes for digital broadcasting in SFN},'' in
	\emph{Proc. 2011 IEEE Workshop Signal Process. Syst. (SiPS)}, Beirut,
	Lebanon, Oct. 2011, pp. 234--238.
	
	\bibitem{6333071}
	A.~{Rico-Alvariño} and C.~{Mosquera}, ``{On the effect of echoes in hybrid
		terrestrial-satellite single frequency networks: Analysis and
		countermeasures},'' in \emph{Proc. 2012 6th Advanced Satell. Multimedia Syst.
		Conf. (ASMS) \& 12th Signal Process. Space Commun. Workshop (SPSC)}, Baiona,
	Spain, Sep. 2012, pp. 168--175.
	
	\bibitem{1189162}
	B.~S. {Yeo} and L.~F. {Turner}, ``{A ratio-based dynamic bandwidth allocation
		algorithm for hybrid LEO satellite and terrestrial networks},'' in
	\emph{Proc. 2002 IEEE Global Telecommun. Conf. (GLOBECOM)}, vol.~3, Taiwan,
	China, Nov. 2002, pp. 2915--2919.
	
	\bibitem{n9}
	I.~F. {Akyildiz}, {J. Xie}, and S.~{Mohanty}, ``{A survey of mobility
		management in next-generation all-IP-based wireless systems},'' \emph{IEEE
		Wireless Commun.}, vol.~11, no.~4, pp. 16--28, Aug. 2004.
	
	\bibitem{5956545}
	M.~{Crosnier}, F.~{Planchou}, R.~{Dhaou} \emph{et~al.}, ``Handover management
	optimization for {LTE} terrestrial network with satellite backhaul,'' in
	\emph{Proc. 2011 IEEE Veh. Tech. Conf. (VTC)}, Yokohama, Japan, May. 2011,
	pp. 1--5.
	
	\bibitem{p255}
	D.~{Liu}, C.~{Huang}, X.~{Chen} \emph{et~al.}, ``{Space-terrestrial integrated
		mobility management via named data networking},'' \emph{Tsinghua Sci. Tech.},
	vol.~23, no.~4, pp. 431--439, Aug. 2018.
	
	\bibitem{6364249}
	M.~{Sadek} and S.~{Aïssa}, ``{Handoff algorithm for mobile satellite systems
		with ancillary terrestrial component},'' in \emph{Proc. 2012 IEEE Int. Conf.
		Commun. (ICC)}, Ottawa, ON, Canada, Jun. 2012, pp. 2763--2767.
	
	\bibitem{7510941}
	G.~N. {Kamga}, M.~{Sadek}, and S.~{Aïssa}, ``{Adaptive handoff for
		multi-antenna mobile satellite systems with ancillary terrestrial
		component},'' in \emph{Proc. 2016 IEEE Intern. Conf. Commun. (ICC)}, Kuala
	Lumpur, Malaysia, May 2016, pp. 1--6.
	
	\bibitem{n10}
	L.~{Fan}, M.~E. {Woodard}, and J.~G. {Gardiner}, ``{Architecture and protocols
		in an IP-based integrated terrestrial/satellite mobile communications
		network},'' in \emph{Proc. 2001 IEEE Int. Conf. Commun. (ICC)}, vol.~9,
	Helsinki, Finland, Jun. 2001, pp. 2850--2854.
	
	\bibitem{n22}
	M.~{Giordani} and M.~{Zorzi}, ``Non-terrestrial networks in the {6G} era:
	{Challenges} and opportunities,'' \emph{IEEE Netw.}, pp. 12--19, 2020, {Early
		Access}.
	
\end{thebibliography}
\end{document}